\documentclass[prb, preprint, bibnotes, nofootinbib]{revtex4-1}

\usepackage{ wrapfig, graphicx, caption}
\usepackage[hidelinks]{hyperref}
\usepackage{mathtools, float, anysize, amsmath, amssymb, gensymb, amsfonts}
\usepackage{tikz,hyperref, accents, changepage}
\usepackage{subfig}
\usepackage{soul}
\usepackage[normalem]{ulem}
 
\usepackage{comment}

\usepackage{multibib}

\usepackage{colortbl}
\usepackage{xcolor}

\newcounter{anexo}[section]

 \renewcommand{\theanexo}{\Alph{anexo}}
\newcounter{apendice}[section]

 \renewcommand{\theapendice}{\Alph{apendice}}

\begin{document}

\title{Experimentation on stochastic trajectories: from Brownian motion to inertial confined dynamics}

\author{Azul Mar\'{\i}a Brigante}
\author{Corina R\'evora}
\author{Gabriel Fernando Volonnino}
\author{Marcos Dami\'an Perez}

\affiliation{Universidad de Buenos Aires, FCEyN, Departamento de F\'{\i}sica. Buenos Aires, Argentina}

\author{Gabriela Pasquini}
\author{Mar\'{\i}a Gabriela Capeluto}

\affiliation{Universidad de Buenos Aires, FCEyN, Departamento de F\'{\i}sica. Buenos Aires, Argentina}

\affiliation{CONICET-Universidad de Buenos Aires, IFIBA, Buenos Aires, Argentina.}

\date{\today}

\begin{abstract}

Statistical physics courses typically employ abstract language that describes objects too small to be seen, making the topic challenging for students to understand. In this work, we introduce a simple experiment that allows conceptualizing some of the underlying ideas of stochastic processes through direct experimentation. Students analyze stochastic trajectories of beads in a bouncing bed of smaller beads subjected to an external periodic drive. The analysis of the trajectories involves the application of a vast toolkit of statistical estimators that are useful in many fields of physics. 

The following article has been accepted by the American Journal of Physics. After it is published, it will be found at \href{https://pubs.aip.org/aapt/ajp}{\textcolor{blue}{Link}}.
\end{abstract}

\maketitle

\section{Introduction}\label{sec:introduccion}

Stochastic processes encompass essentially everything we speak about in everyday life: the number of customers in a checkout line, the stock market, exchange rate fluctuations, gambling, commuting from home to the workplace, blood pressure and temperature, and bacterial growth, among many others.\cite{Acioli2020, Germain2016, patriarca2013, stauffer2008}. 
Brownian motion, the most well-known type of random (stochastic) particle trajectory, was first described in 1827 by the botanist Robert Brown, who observed the irregular motion of pollen grains suspended in water.\cite{thelangevinEquation,MolecularModelingandSimulation} The explanation of this motion was provided nearly a century later by Einstein in 1905 using the kinetic theory of thermally excited molecules colliding with the particle, providing strong evidence for the existence of atoms and molecules.\cite{Haw_2002} As will be shown in Section \ref{sec:fundamentals}, Brownian motion occurs for a particle in a viscous medium with an external stochastic force (Langevin's equation) in the particular case where the inertial force $F_{\textrm{inertia}}$ is negligible in comparison with the viscosity $F_{\textrm{viscous}}$ and boundary conditions are not relevant. The relative importance of the two forces is captured by the Reynolds number
\begin{equation}
   R\sim \frac{F_{\textrm{inertia}}}{F_{\textrm{viscous}}} \sim \frac{\rho a \langle v \rangle}{\eta},
  \label{eq.reynold}
\end{equation}
\noindent where  $\langle v \rangle$ is the average particle speed, $\rho$ is the particle density, $a$ is the particle diameter, and $\eta$ is the viscosity of the medium.\cite{Purcell}

Particle trajectories subjected to random fluctuations occur in many areas of physics, biology, and chemistry. For example, stochastic trajectories are the underlying processes in classical diffusion between regions with different known concentrations. These concepts are also fundamental for understanding the boundary between two phases in an alloy, the dispersion of contaminants in gases and liquids, and the dynamics of the ionic motion across the cell membranes through single biological channels made of protein molecules.\cite{diffAsChRe, Poydenot2022, Diffspread, cell} Moreover, the predictions from Brownian motion are key to characterizing optical tweezers, which are essential tools to study forces acting between micrometer-scale objects including cells and proteins.\cite{PhysRevLett.24.156}

In this article, we present a simple and inexpensive experiment to study stochastic trajectories that we have successfully introduced in an advanced undergraduate course. The experiment consists of tracking the position of a bead that collides with other smaller beads. To make the collisions random, a speaker shakes the small beads. Due to the specific choice of experimental configuration (see section \ref{sec:setup}), the hypotheses related to Brownian dynamics will be challenged and rigorously tested, contributing significantly to a deeper understanding of concepts associated with stochastic processes. Specifically, Brownian motion requires that inertia is negligible compared to friction and that the random force is Gaussian and uncorrelated. In this experiment, students will be prompted to test each hypothesis against their findings.

This experience will enable students to conceptualize out-of-equilibrium phenomena through experimentation. In doing so, they will acquire knowledge of statistical methods applicable across a wide range of natural scenarios where stochastic processes emerge. Additionally, since this implementation involves tracking particle positions from images captured by a CCD camera (stacked in videos), students will also gain proficiency in using image processing tools. Students at this level are expected to have a solid foundation in mechanics and thermodynamics, along with some basic concepts of statistics.

The article is organized as follows: In Section \ref{sec:fundamentals}, the theoretical fundamentals of Langevin dynamics and Brownian motion are reviewed. In Section \ref{sec:setup} we describe the experimental setup. In Section \ref{sec:results and discussion} we present an example of results obtained by a group of students during three consecutive classes (6 hours each) of a full semester experimental course, during which they developed the tracking code and collected the data. The statistical tools necessary to perform the data analysis are introduced. Finally, in Section \ref{sec:conclusions} we share some concluding remarks.

\section{Probabalistic trajectories}\label{sec:fundamentals}

\subsection{Langevin formulation}

Let's start with a particle of mass $m$ and diameter $a$ that is subjected to a deterministic force $f_{\text{det}}$ in a medium with viscosity $\eta$, at temperature $T$. The thermal motion of the molecules in the surrounding medium produces impacts on the particle resulting in impulsive forces. If the particle is large compared to the molecules but small enough that the effect of the random collisions with multiple molecules in a short time is non-negligible, a fluctuating force $f_{\text{rand}}(t)$ must be included in Newton's equation: 
 \begin{align}
     m\ddot{x}=-\gamma  \dot{x}+ f_{\text{det}} + f_{\text{rand}},
     \label{ec:langevin}
 \end{align}
where $\gamma = 6\pi \eta a$ is the drag coefficient.\cite{thelangevinEquation} The random force is assumed to be independent of the position and its variations in time are much faster than the variations of $x(t)$, in such a way that at a given position,  ${\langle f_{\text{rand}}\rangle} = 0$. This equation, known today as Langevin's equation, was first proposed in 1908.\cite{thelangevinEquation} In this formulation, the random force is assumed to be uncorrelated, so that its auto-correlation in time is
\begin{equation}
        \langle f_{\text{rand}}(t+\tau) f_{\text{rand}}(\tau) \rangle_{\tau} = \int_{-\infty}^{\infty}f_{\text{rand}}(t+\tau) f_{\text{rand}}(\tau) d\tau 
        \sim \delta(t),
        \label{ec:correlation}
\end{equation}
with $\delta(t)$ the Dirac delta function. 

Let's examine now the relation between the auto-correlation and the size of the fluctuations, as characterized by the variance $\sigma_f^2=\langle (f_{\text{rand}} -\langle f_{\text{rand}}\rangle)^2 \rangle$. By setting $t=0$, the autocorrelation simplifies to the mean square random force $\langle f_{\text{rand}}^2 \rangle$, which, for a random force that averages to zero ($\langle f_{\text{rand}} \rangle$=0), is equal to the variance ($\sigma_f^2=\langle f_{\text{rand}}^2 \rangle$). Therefore, the variance for an uncorrelated random force, representing the size of the fluctuations, is a constant independent of time and frequency, that is, a white noise. In the case that $f_{\text{det}}=0$, Langevin's equation can be solved to determine the average squared velocity $\langle v(t)^2 \rangle$, which can be related to the random force as $\langle f_{\text{rand}}^2 \rangle = 2 \gamma m \langle v^2\rangle$ in the limit of large time. \cite{pathria} Since the system is expected to reach thermodynamic equilibrium at this point, we can use the equipartition theorem $\langle v^2\rangle = k_B T/m$ to express this equation as $\langle f_{\text{rand}}^2 \rangle = 2\gamma k_B T$, with $k_B$ the Boltzmann constant. This expression is a manifestation of the fluctuation-dissipation theorem,  which relates the correlation of the random force with the viscosity and explains how the energy gained by the particles through collisions with other particles in the medium is quickly dissipated in the same medium by friction.\cite{thelangevinEquation}

Since a large number of molecules are colliding with the particle, there will be a large sample of random independent forces. Then, according to the central limit theorem, the probability density function (PDF) for $f_{\text{rand}}$ must approach a normal distribution centered at zero with variance $\sigma_f^2$.  

In order to numerically simulate the possible trajectories, a discrete expression can be found from Eq.~\ref{ec:langevin}:
\begin{subequations}
\begin{align}
    x_{n+1} &= x_n + v_n \Delta t\\ 
        v_{n+1} &= v_{n} -\frac{\gamma}{m} v_n \Delta t -\frac{f_{\text{det}}}{m}\Delta t + \Delta v_{\text{rand},n}   ,
\end{align}
\label{ec:sim_Langevindiscr}
\end{subequations}
where  $ v_n =\dot{x}_n=\Delta x_n/\Delta t $ is the velocity in the step $n$ and $\Delta v_{\text{rand},n}$ is the random change in velocity due to the average random force in this step. Note that, while a Runge-Kutta method would surely give better results in the simulated trajectories at large time scales, these considerations go beyond the scope of the present undergraduate experimental proposal and, as will be shown, the obtained results are quite reasonable using this simpler method.

\subsection{Brownian regime} \label{sec:browinian}

The simplest case of Brownian dynamics emerges in the \textit{overdamped} limit of Langevin's equation when the inertial term can be ignored; that is, when   $ m\ddot{x}\ll \gamma \dot{x} $ in Eq. ~\ref{ec:langevin}. In this case, by approximating the velocity by a first-order numerical expression ($\dot{x} \approx \Delta x/ \Delta t$) Eq.~\ref{ec:langevin} can be reduced to a single discrete equation\cite{AGentleIntroduccion}

\begin{align}
   \Delta x = f_{\text{det}}\frac{\Delta t}{\gamma} + \Delta x_{\text{rand}} =\Delta x_{\text{det}}+\Delta{x_{rand}}.
    \label{ec:overdamp_lang}
\end{align}

Here $\Delta x_{\text{det}}$ is a deterministic step due to the deterministic forces and $\Delta x_{\text{rand}}$ is the random contribution to the particle's path. 

Random motion may also occur in the absence of deterministic forces. Einstein arrived at a solution for this problem in 1905 by showing that the Brownian PDF follows a diffusion equation. He considered the simplest form of a random walk, involving out-of-equilibrium displacements, and made the assumption that the velocity distribution follows the equilibrium Maxwell-Boltzmann distribution. \cite{EinsteinPaper, thelangevinEquation} By using the same concepts as the fluctuation-dissipation theorem, he considered that collisions only produce random jumps in the positions of the particles. In his theory, Einstein showed that, in the limit of very small steps, the PDF follows a diffusion equation. \cite{EinsteinPaper, thelangevinEquation} In the absence of deterministic forces, the PDF for $\Delta x$ is a Gaussian distribution 

\begin{equation}
    p_1(\Delta x)=\frac{1}{\sqrt{2 \pi} \sigma}e^{-\Delta x^2/2\sigma ^2}
    \label{ec:SSprob}
\end{equation}
with variance $\sigma^2 = 2 m \langle v^2 \rangle \Delta t/\gamma = 2k_B T\Delta t/\gamma$, because particles are in thermal equilibrium following a Maxwell-Boltzmann distribution, and therefore the equipartition theorem holds.   
 
\begin{figure}[h]
    \centering
    \hspace*{-.3 cm}\includegraphics[scale=0.48]{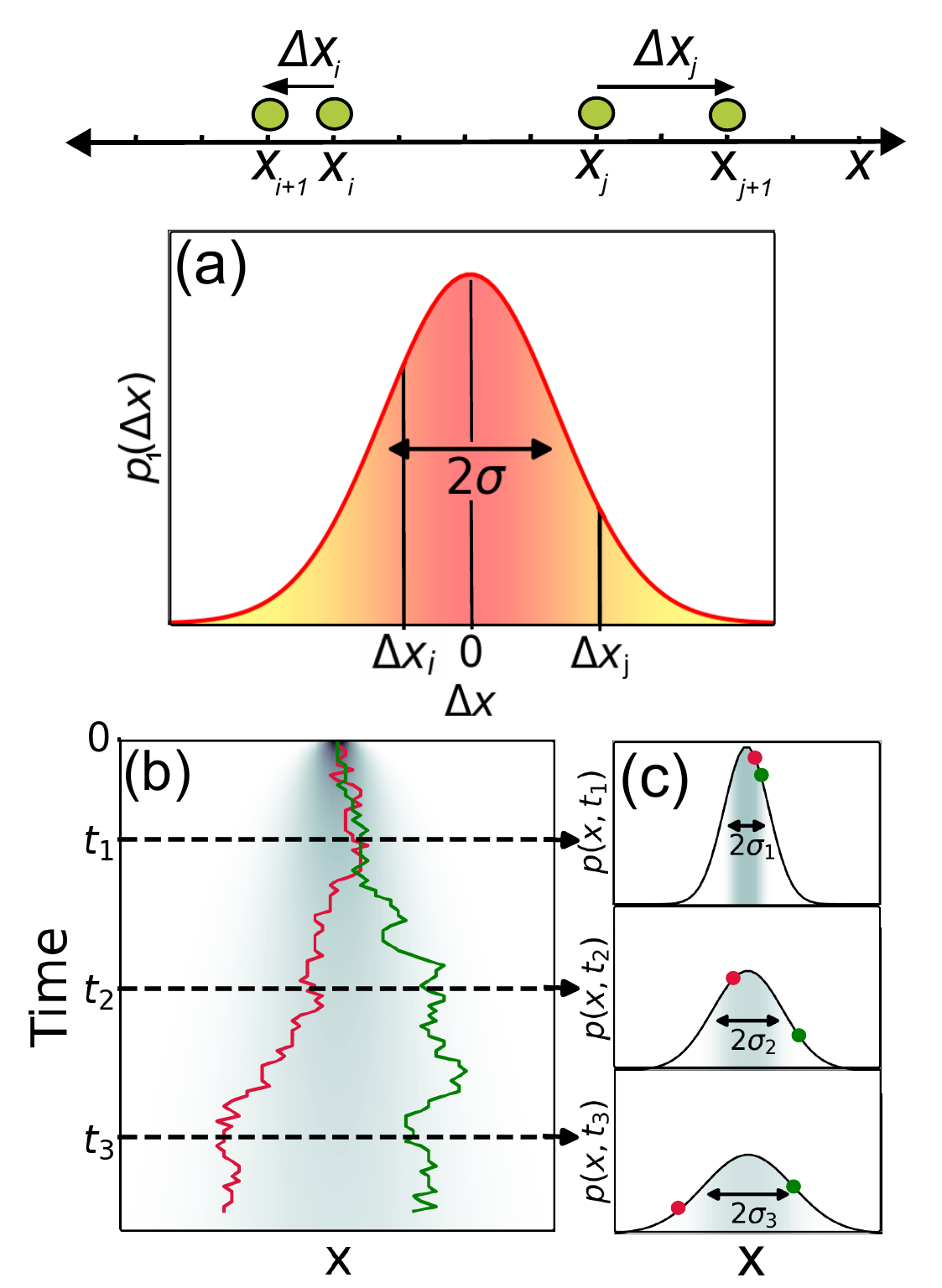} 
    \caption{(Color online) (a) Schematic of the Gaussian PDF, $p_1(\Delta x)$, for displacements $\Delta x$ at fixed time ($t$), with variance $\sigma^2 (t)$. The scale on top shows possible positions for a particle at time $t$ and a particular example (black dots) with its corresponding probability.
    (b) Examples of two trajectories (red and green curves) simulated using eq. \ref{ec:sim1} and (c) the theoretical PDF from eq. \ref{ec:pos_distr}, $p(x,t)$, for selected times ($t_1$, $t_2$ and $t_3$). The double arrows indicate the width of the PDF measured when the Gaussian reduces to $e^{-1/2} \sim 0.6$ from its maximum value, which gives $2 \sigma$. Another common criterion is the Full-Width at Half Maximum, from which the width results in $ \sim 2.35 \sigma$.  }
    \label{fig:pdf_displacmen_intro}
\end{figure}

Fig.~\ref{fig:pdf_displacmen_intro}(a) shows a scheme of this distribution with its relevant parameters. We use the notation $p_1(\Delta x)$ for this probability distribution of step sizes in order to distinguish it from the probability distribution $p (x)$ that we will introduce in the next section to describe entire trajectories.

A numerical implementation of Eq.~\ref{ec:overdamp_lang} can be derived by defining $t_n=n\Delta t$ and $x_n= x(t_n)$ with initial condition $x(0)=x_0$, which leads to the simplest equation able to simulate stochastic trajectories 
\begin{equation}
    x_{n+1}=x_n + \Delta x_n,
    \label{ec:sim1}
\end{equation}
where $\Delta x_n$ are random numbers taken from the distribution $\Delta x_\text{rand}$. Figure \ref{fig:pdf_displacmen_intro}(b) shows examples of simulated trajectories (red and green curves). As is evident, particles originating from the same starting position can ultimately become widely dispersed. As it will be shown in section \ref{proba}, the position at fixed times follows a Gaussian distribution centered at the average position (figure\ref{fig:pdf_displacmen_intro}(b)). The shaded gray area illustrates the expected increase in standard deviation over time, which will be quantitatively derived in the next section.

\subsection{Langevin oscillatory dynamics}

In experiments, it is likely that, in addition to random thermal forces, other types of attractive or repulsive forces may arise. Moreover, the assumption that the inertial term can be neglected may not necessarily be valid.  To model inertial dynamics in an attractive potential (that may represent confinement), the simplest deterministic force that can be introduced into the Langevin equation is the restoring force given by $f_{\text{det}} = -m \omega_0^2 x$, where $\omega_0^2 = k/m$ and $k$ is a stiffness constant. 
 
This kind of restoring force can be also used to model an optical trap or optical tweezers. By including it in Eq.~\ref{ec:sim_Langevindiscr},  we arrive to the following set of discrete equations:\cite{Ren11}
\begin{subequations}
\begin{align}
    x_{n+1} &= x_n + v_n \Delta t\\ 
    v_{n+1} &= v_{n} -\beta v_n \Delta t -\omega_0^2 x_n \Delta t  + \Delta v_{\text{rand},n} 
\end{align}
\label{ec:sim_LMCP}
\end{subequations}
\noindent
where $\beta=\gamma/m$ is the drag coefficient associated with the viscosity and $\Delta v_{\text{rand},n}$ is a random number obtained as in Ref.~\onlinecite{Ren11} that follows a Gaussian distribution with standard deviation $\sigma = \sqrt{2 k_B T \gamma \Delta T/m^2} = \sqrt{2\langle v^2\rangle \beta \Delta t}$. Now the PDF for the displacements will be centered at each time around a deterministic displacement $\Delta x_\text{det}(t)$. 

\section{Experiment} \label{sec:setup}

\begin{figure}[htp]
    \centering
    \includegraphics[scale=0.7]{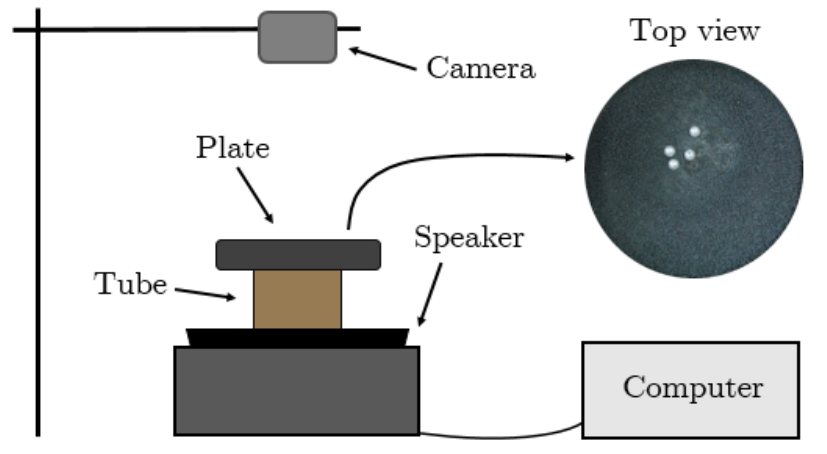} 
    \caption{Diagram of the experimental setup. A speaker controlled by a computer is used to shake a plate with beads. The plastic beads (white) and steel beads (gray), are observed in the top view of the plate.  Different signals can be programmed in the computer in order to produce stochastic trajectories. The trajectories were obtained from videos acquired by a camera ($30\,\mathrm{fps}$).}
    \label{fig:setup}
\end{figure}

The experimental setup designed to produce stochastic trajectories is shown in Fig.~\ref{fig:setup}. It consists of a bed of bouncing steel beads (diameter 0.9$\pm$0.1~mm and weight $3.8\pm 0.2$~mg) driven by a speaker and recorded with a CCD webcam. A picture of the beads in the experimental setup is shown in the top view in Figure \ref{fig:setup}. The beads are placed in a 12 cm diameter and 2 cm depth dish that is connected to a 6" speaker diaphragm with a cardboard tube. The speaker (6 $\Omega$ impedance, 60 ${\rm W}_{{\rm RMS}}$ maximum) is driven by a filter-free class-D Stereo Amplifier (TPA3116D2 from Texas Instruments) powered with a 24 V switching source to provide 30 ${\rm W}_{{\rm RMS}}$ maximum power. The amplifier is connected to the computer, which allows the production of sounds with arbitrary amplitudes and frequencies with simple commands, as shown in the supplementary material. \cite{supplementary}  

The experiment consists of tracking the 2D trajectories of a larger plastic bead (diameter 6.00 mm and weight 103.8 mg) placed on top of the steel beads while they are being shaken by the speaker. The steel beads were obtained from a steel shot blasting media that can be purchased in a hardware store and the white plastic bead is just a pearl bead from a necklace. The steel beads impact the plastic bead producing random impulsive forces, and therefore the plastic bead's trajectories are stochastic. The specific choice for the proposed experimental design is quite different from the standard Brownian experiment. In this experiment,  the particle is only 6 times larger and 25 times heavier than the particles that cause the random forces. However, typical experiments studying Brownian motion utilize particles that are at least three orders of magnitude larger than the particles in the environment and very much heavier. \cite{Darras2017, Stephan2023, Poydenot2022, Nakroshis2003, Acioli2020, Germain2016, Krishnatreya2014} Additionally, the particle's movement is limited to a plate that is only 20 times larger than its diameter. As will be shown shortly, it is important for the students to be aware of these conditions when analyzing the data, since this particular selection challenges the hypothesis of a pure Brownian motion. This opens up possibilities for a rich discussion on physics and challenges the students towards critical thinking \cite{youtube1,youtube2}. In this direction, the main questions we ask students to address through data analysis are oriented toward analyzing the hypothesis that assumes pure Brownian motion. These questions include: can the motion of these beads be described as Brownian? Can the inertial term be effectively neglected? Does confinement affect the trajectories? Is the hypothesis of random noise fulfilled?

The plate that holds the particles is essentially a Chladni plate that will oscillate at higher amplitudes at its resonance frequencies. However, when the plate oscillates in a single normal mode, synchronized oscillations induce correlations in the trajectories. Furthermore, in normal modes, there are large areas of zero amplitude (nodes) that should be avoided so that the particles are not trapped. On the other hand, in order to be able to observe the trajectories for long enough to obtain good statistics, it is necessary to ensure that beads remain moving for a long enough time without reaching the edges of the plate. Therefore, to ensure the production of stochastic collisions in a confined region, the signal sent to the speaker must be carefully selected. For this reason, an initial characterization of the oscillation modes was made by applying sinusoidal signals $A_{\nu} \sin(2\pi \nu t)$ with frequencies $\nu$ in the audible spectrum (20 Hz to 20 kHz) and obtaining the frequencies that were seen by eye to maximize the oscillations. We then programmed several superpositions of normal modes, and we looked for the frequency combination that successfully prevents trajectories from becoming correlated: This means the particle jumps from one position to another unpredictably. If the movement of the particle synchronizes with the oscillation of the dish, then the trajectory may be predictable, and the steps and trajectories could be correlated. Despite this, there is a great playground for exploration in terms of plate oscillations. For example, students are welcome to play with different signals and analyze the implications in the trajectories.   

Finally, the trajectories of the plastic beads are obtained from videos acquired using the CCD camera at 30~fps. As the sampling frequency is much smaller than the mean collision frequency (which is at least the higher frequency at which the speaker vibrates), there is another source of randomization given by the process of sampling. The color of the beads was chosen in order to have high contrast in the images, which allows for improving the performance of the tracking algorithm. The videos are acquired in RGB colors and converted to grayscale levels. Each video is composed of a stack of $N$ images with $1280 \times 720$ pixels. Therefore the information is stored in a 3-D indexed matrix $V(i,j,k)$, with $i, j$ belonging to each frame (each image) and $k$ enumerating the frames (i.e. the evolution in time). The spatial scale of the frames is previously calibrated, resulting in an interpixel distance of $(0.285 \pm 0.002)$ mm. Finally, the time between consecutive frames (i.e. the inverse of the frame rate, 1/30 fps), is taken as the time step of the trajectories $\Delta t$. 

Many image processing techniques for pattern recognition, for example, Template Matching or Colocalization by cross-correlations are useful for tracking algorithms. Utilizing these techniques, our students implemented a simple tracking code in Python, based on cross-correlations. In cases of limited time, free software like Tracker can be used. However, we believe it is valuable for students to develop their own programs instead of relying on commercial ``black box'' solutions. The general idea is to find, in each frame, a region similar to a template that represents an image of one plastic bead. The algorithm then searches for the region in each frame that maximizes the cross-correlation with the template and records its position. While it's possible to film multiple particles simultaneously, the tracking code will only track a single particle at a time.. The flow diagram of the tracking code, along with specific numerical details and examples, are included in the supplementary material.\cite{supplementary}

\section{Results, analysis, and discussion}\label{sec:results and discussion}

\begin{figure}[h]
    \centering
    \includegraphics[scale=0.3]{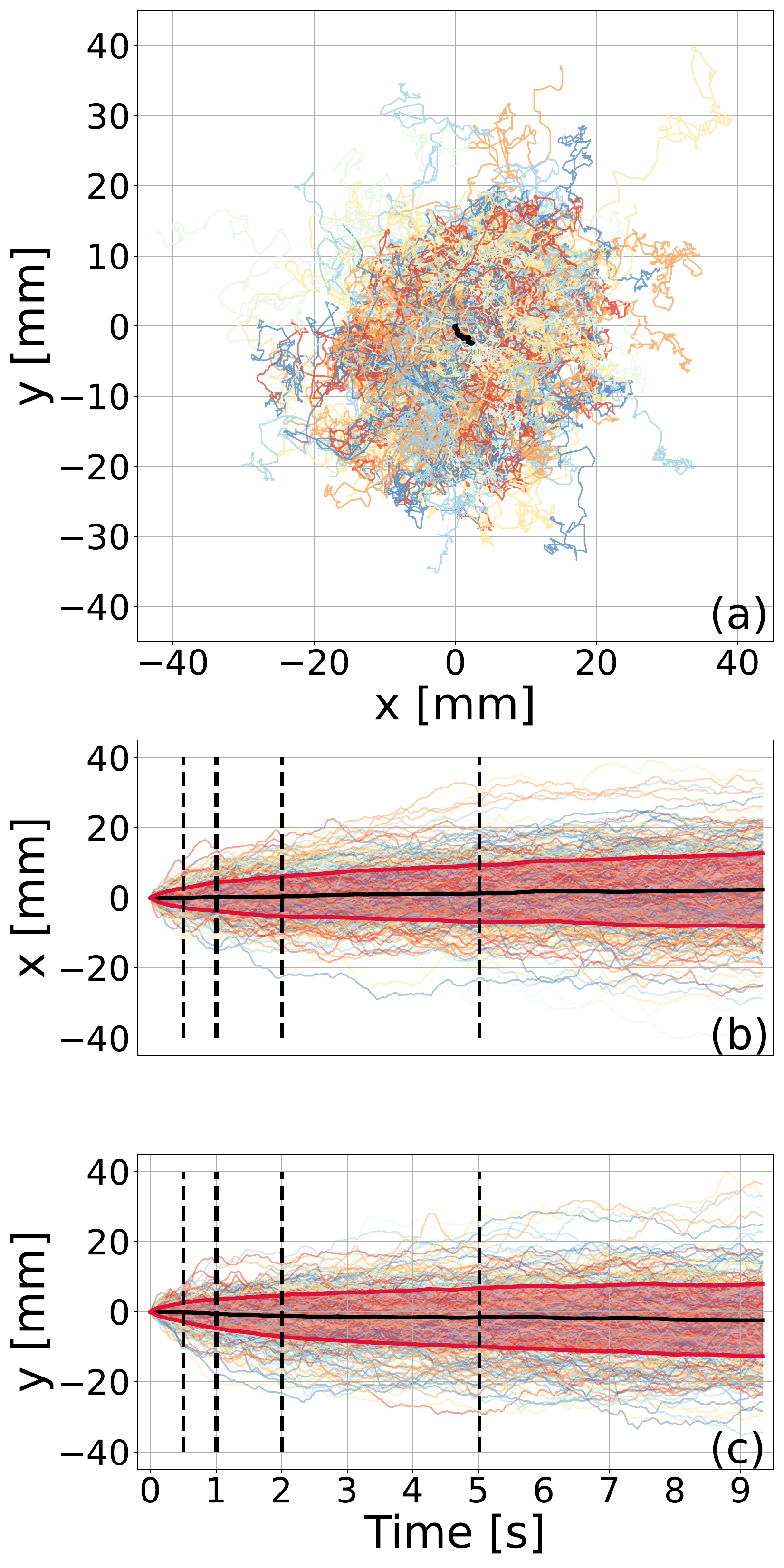} 
    \caption{(a) Measured 2D trajectories for the white beads. (b) and (c) x and y coordinates of the measured trajectories (color lines) together with the average trajectory (black line) and the standard deviation (red shadow). Vertical black dashed lines indicate the times at which the histograms from Figure \ref{fig:dist_x_vertical} were calculated.}
    \label{fig:trayectories}
\end{figure}

Figure \ref{fig:trayectories}(a) shows about 500 measured trajectories in the $x-y$ plane, with a common origin at $(0,0)$. The average 2D trajectory is also plotted with a thicker black line, evidencing a small average drift. The time dependence for the $x$ and $y$ coordinates of the trajectories are shown in panels (b) and (c) respectively. This figure illustrates important features of stochastic processes. On the one hand, measuring the position of a particle over time results in a random path as the outcome of the stochastic process. The set of data points giving dependence of the position variable on time is termed a trajectory (illustrated by the colored lines in Fig 3a and b). Every trajectory is a random function. On the other hand, when measuring positions at a fixed time (along the vertical black dashed lines in Fig. 3b and 3c), the outcome is a random variable. The collection of all trajectories is known as an ensemble. Throughout the rest of the article, certain calculations will be performed on the random variables, while others will be conducted on the random trajectories.

It can be seen that the average trajectories in both x and y directions (black lines in Figure \ref{fig:trayectories}~(b) and (c)) drift linearly in time. This may be due to a slight inclination of the bead holder that could lead to a small projection of the gravity force in the plane of motion (i.e. a constant deterministic force in  Equation \ref{ec:overdamp_lang}). It is easy to see that, if the overdamped approximation holds, this force leads to an average position linear with time. A drift velocity for each coordinate can be obtained from the slope of a linear regression on the average trajectories, which was found to be ($0.250\pm0.003$)~mm/s and ($-0.214 \pm 0.005$)~mm/s for $x$ and $y$, respectively. 

Besides this drift, it can be seen by eye that the dispersion of the trajectories increases with time, as is expected for a stochastic process.  The dispersion can be quantified by computing the standard deviation (red lines in Figure \ref{fig:trayectories}~(b) and (c)). 

In the following subsection, we focus on the statistical analysis of the time evolution and correlations of the measured trajectories by comparing their properties with those expected from the solutions of the different motion equations presented in Sec.~\ref{sec:fundamentals}. Complementary data analysis is proposed in the supplementary material.\cite{supplementary}
 
\subsection{Probability density function (PDF)\label{proba}}

As mentioned above, the trajectories exhibit a tendency to disperse significantly over time. Since the precise trajectory is clearly unpredictable, it is attainable to analyze the average properties of the several possible paths, for which the trajectory's probability needs to be calculated. For simplicity, we show the results of the calculations in one dimension. The probability density function (PDF) for an entire given trajectory is the product of the probabilities $p_1(\Delta x_j)$  found in Eq.~\ref{ec:SSprob} for each step $\Delta x_j$ (joint probability): $\prod_{j=1}^N p_1(\Delta x_j)$, where the steps are considered to be independent\cite{AGentleIntroduccion}. One can then find the total probability for all the trajectories that arrive at the same endpoint $x$ in a time $t$, which for differentially small steps becomes (see for example Refs [\citenum{AGentleIntroduccion, Lemons2002, thelangevinEquation}]):

\begin{align}
    p(x, t)= \frac{1}{\sqrt{2\pi} \sigma(t)} \exp{\left[-\frac{(x-\Bar{x})^2}{2\sigma^2(t)}\right]},
    \label{ec:pos_distr}
\end{align}
where $\sigma^2(t) = 2k_B T t/\gamma$ and $\Bar{x}$ is the position corresponding to the average trajectory at time t, or simply the average trajectory. Note that the probability distribution becomes broader and flatter over time and is centered at $\bar{x}$, as shown in Figure \ref{fig:pdf_displacmen_intro}(c). 
 
A PDF is characterized by the expectation values of $(x-\bar{x})^m$,  with $m$ and integer number, which are also called the central moments of the distribution:
\begin{equation}
    \mu_m=E[(x-\bar{x})^m]=\int_{-\infty}^{\infty} (x-\bar{x})^m p(x,t)dx.
    \label{ec:moments}
\end{equation}

With this definition, the first central moment $\mu_1=E[(x-\bar{x})]=0$.
 
The second moment $(m=2)$ of a normal distribution is the variance $E[(x-\bar{x})^2]=\langle (x-\bar{x})^2 \rangle = \sigma^2(t)$. The standard deviation $\sigma(t)$ (squared root of the variance) expected in Brownian dynamics for different times, is shown as a gray shadow in Fig.~\ref{fig:pdf_displacmen_intro}(b) and (c). 

\begin{figure}[htp]
    \centering
    \includegraphics[scale=0.39]{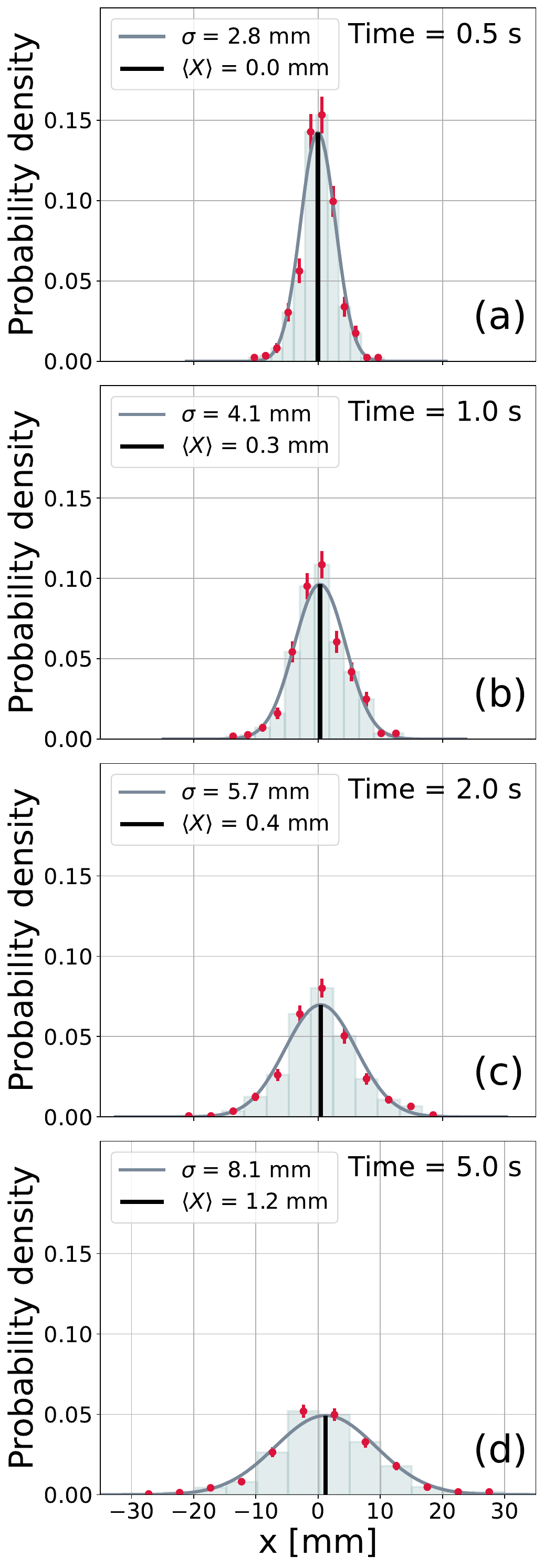} 
    \caption{Histograms for the positions at fixed times (as shown in Figure (\ref{fig:trayectories})) for the x-coordinate of the trajectories, and the corresponding Gaussian curves (indicated by the black lines). The bin sizes of the histograms are 1.8 mm(a), 2.3 mm (b), 3.6 mm (c), and 5.0 mm (d). For each time, the values of the average trajectory $\langle x \rangle $ and the standard deviation $\sigma$ computed from the data (shown in each panel) were used to plot the PDFs.}
    \label{fig:dist_x_vertical}
\end{figure}

We can experimentally estimate the probability of finding a particle at a given position and time from the histograms of the positions of a bead in each trajectory at fixed times. We arbitrarily choose specific times in Fig.\ref{fig:trayectories} (a) and (b) (shown as black dashed lines) and construct the histograms shown in Fig.\ref{fig:dist_x_vertical}. These histograms are normalized such that the integral of the probability density over distances (expressed in mm) equals unity. Because the distribution of counts in a histogram bin is binomial, the error $\epsilon_{i}$ of a bin of height $p_i = N_i/N$ is estimated as the binomial error $\epsilon_{i} = (p_i \, q_i /N)^{1/2} $,  where $N_i$ is the number of entries in the bin $i$, $N$ is the number of total entries in the histogram and  $q_i = 1-p_i$ \cite{Bevington}. The error bars of the probability density (red vertical lines in the figure) are then estimated from the error of the probability of each bin with further normalization.

The definitions given in equation \ref{ec:moments} for the central moments of a given PDF  also apply to the central moments of a discrete sample distribution. Therefore, we can numerically compute the central moments (for example, the variance) associated with the measured trajectories from our data set at each time. In Figure \ref{fig:dist_x_vertical}  we show the Gaussian distribution obtained using the variance computed from the trajectories at the previously selected times. A deviation of the data distribution from the Gaussian statistics would suggest that the motion is not strictly Brownian. Exploring this potential deviation can be a valuable approach to teaching data analysis. This is not the primary focus of our proposed study, but we include the topic in the supplementary material. \cite{supplementary}

A more relevant study for stochastic processes involves investigating how the variance and other characteristic observables, which will be defined shortly, spread over time. For example, in the case of Brownian dynamics, it is expected that the variance will increase linearly with time. Let's examine the time evolution of our experimental data.

\subsection{MSD evolution and diffusion}

The fact that the probability distribution gets flatter and broader as time increases is a signature of diffusive behavior. The observable typically used to study the dispersion of the trajectories is the variance. In strong connection, the observable describing the diffusion of the particles in time is the mean square displacement (MSD)
\begin{align}
    \text{MSD}(t)=\langle(x(t)-x_0)^2 + (y(t)-y_0)^2 \rangle , 
    \label{ec:MSD}
\end{align}
that quantifies the deviation of a particle position with respect to a reference position $(x_0, y_0)$ after an elapsed time $t$. In the general case, the reference position can be any position in the trajectory. Under some reasonable assumptions, averaging the position of ``infinite'' possible trajectories at a fixed time $t$ should give the same result as averaging the distance traveled by a particle during a time interval $t$ in a single ``infinite" trajectory. Therefore the MSD coincides with the variance in the infinite limit. However, to emphasize the conceptual difference, variance quantifies the spread or dispersion of positions around the average, whereas MSD describes the average squared distance that a particle has moved from a certain position over a given time interval. In principle, any position can be assumed as the ``initial position". Therefore, the $\text{MSD}(t)$ can be estimated using just one of the trajectories, by averaging, all the distances between points separated by a time interval $t$.

The relevance of the MSD in this kind of stochastic process mainly arises from its relationship with the diffusion coefficient $D$. Einstein found out that Brownian motion is produced by collisions, and arrived at a diffusion equation for the probability distribution given in Equation \ref{ec:pos_distr}\cite{EinsteinPaper} with diffusion coefficient $D = \sigma^2 /2 \Delta t =\langle v^2 \rangle/\beta$ (See Section \ref{sec:browinian}). Therefore, neglecting the inertial term and in the absence of deterministic forces, the MSD grows linearly with time and it is proportional to $D$,
$\text{MSD}(t)=2 D t$. Measurements of the MSD as a function of time can indicate whether other forces are present since they can cause either superdiffusive (faster growth than linear) or subdiffusive (slower growth than linear) behavior. In a general case, the instantaneous diffusion coefficient is defined as $D(t)= \partial_t \text{MSD}(t)/2$. The diffusion coefficient is determined in the long time limit, being calculated as $D = \lim_{t \to \infty} D(t)$. \cite{arxiv.1906.10402, msdArticle}

\begin{figure}[h]
    \centering
    \includegraphics[scale=0.32]{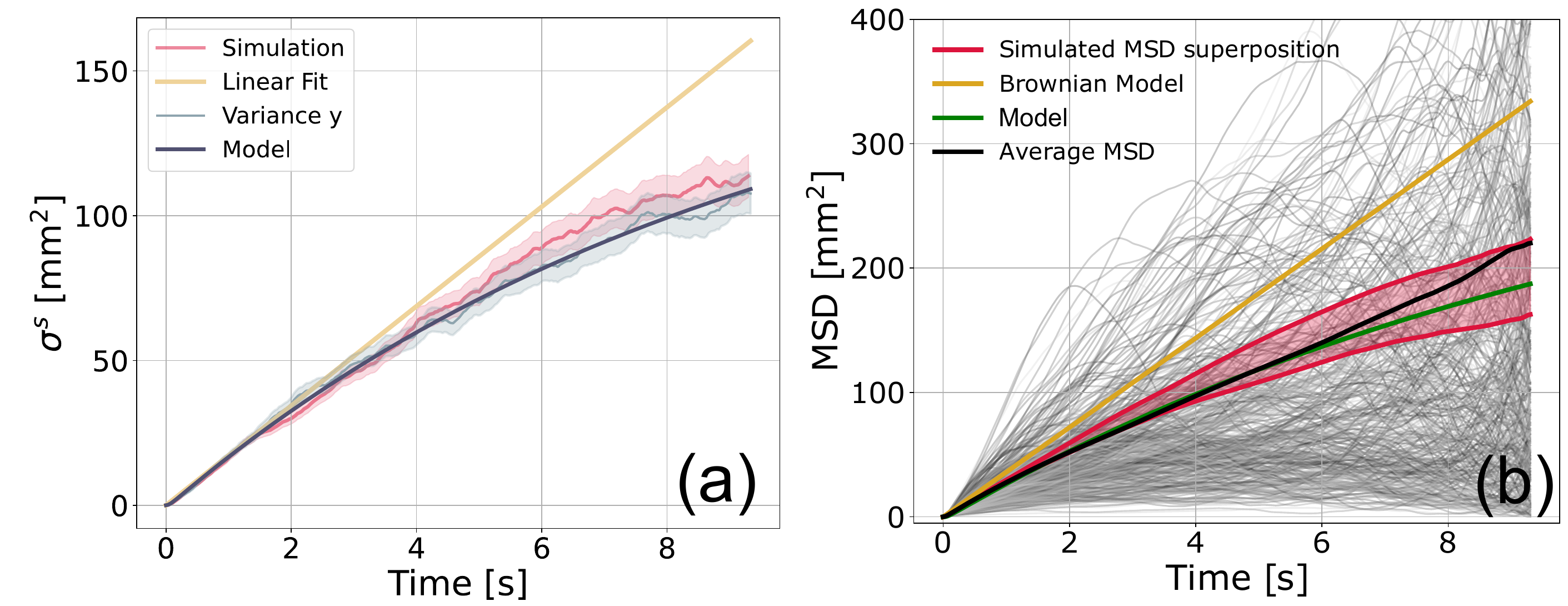} 
    \caption{(Color online) (a) Dependence of the variance with time.
    A linear growth only holds for the first two seconds and is shown with a linear fit during this period (red line). In dark gray is the fitting of variance using the model of an inertial particle in a quadratic potential. In brown are the results of a 500 trajectories simulation with the parameters obtained from the fitting. (b) MSD as a function of the time interval $t$ for each measured trajectory (gray lines) and average experimental MSD (black line). The MSD for a Brownian motion (yellow line) and the MSD model for an overdamped particle (green), both calculated with the experimental parameters, confirm a subdiffusive regime. In red, the superposition of the MSD computed for 500 subgroups of 500 simulated trajectories each (See Supplementary material \cite{supplementary}).}
    \label{fig:varianceandMSD}
\end{figure}

Figure \ref{fig:varianceandMSD}~(a) shows the experimental variance of the $y$-component of the position (the $x$-component behaves similarly) with its error estimated as the standard error $\text{SE}_{\sigma^2} = \sqrt{2 \sigma^4/(N-1)}$, where $N$ is the number of samples (number of trajectories) and $\sigma^2$ its standard deviation.\cite{Ahn03standarderrors} It can be seen from the linear regression that a linear behavior only holds for the data at times shorter than two seconds. From this initial slope of ($16.43\pm0.04$)~m/s$^2$, a diffusion coefficient $D=\sigma^2/2\Delta t $ can be obtained for short times (Table \ref{tab:fitparametersConfined}). For longer times the variance grows more slowly than expected, meaning that the movement is subdiffusive. Three factors related to the experimental design can lead to subdiffusive trajectories: (1) the edges of the plate confine the trajectories to a limited space; (2) although we are utilizing a combination of normal modes with the intention of avoiding nodes, it is still possible to have certain regions with smaller amplitudes that act as confining regions for trajectories; (3) it is not obvious for our experimental conditions, that the inertial term can be neglected. Let's explore these possibilities. 

\begin{figure}[h]
    \centering
    \includegraphics[scale=0.4]{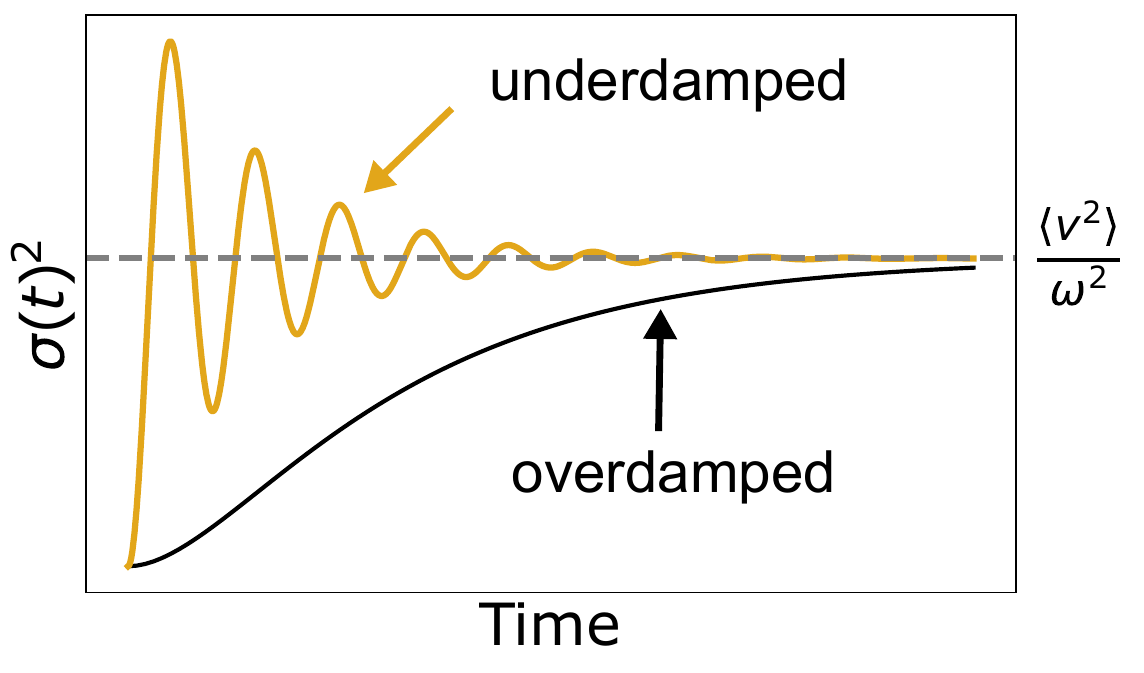} 
    \caption{Schematic of the possible evolution of  variances with time, obtained as solution of Equation \ref{ec:langevin} for an overdamped oscillator (black curve) and an underdamped oscillator (red curve)}
    \label{fig:variance_intro}
\end{figure}

As discussed in Section \ref{sec:fundamentals}, a deterministic harmonic potential provides the simplest model that includes inertia and confinement. The time dependence of the variance associated with the trajectories obtained by solving equation \ref{ec:langevin} with an harmonic deterministic force depends on whether the oscillation is \textit{overdamped} ($\beta > 2\omega_0$, black curve in Figure \ref{fig:variance_intro} ), \textit{underdamped} ($\beta < 2\omega_0$, red curve in Figure \ref{fig:variance_intro}) or \textit{critically damped} ($\beta = 2\omega_0$). In particular, Lemons showed that the dependence of the variance with time for an overdamped oscillator is given by\cite{Lemons2002}\\
\begin{equation} \label{eq:sigmaarmonico}
\begin{split}
    \sigma^2(t) = \frac{\langle v^2\rangle}{\omega_0^2} \left[1 - e^{-\frac{1}{2}\beta t} \left(2\sinh^2{ \frac{1}{4}\alpha t} + \frac{\beta}{\alpha} \sinh{\frac{1}{2}\alpha t} + 1\right)  \right], 
\end{split}
\end{equation}
where $\alpha = \sqrt{\beta-8\omega_0^2}$. The variance at $t=0$ is therefore given by $\sigma^2(0) = A=\langle v^2\rangle/\omega_0^2$.  The MSD and the average trajectory $\langle x(t) \rangle$ in this regime are given by:
\onecolumngrid
\begin{equation} \label{ec:MSDconf}
\begin{split}
    \textrm{MSD}(t) &= \langle  x(t) \rangle ^2 + \frac{\langle v^2\rangle}{\omega_0^2} \Big[ 1  - e^{\alpha' t}\Big(2\frac{\beta^2}{\alpha'^2}
    \cdot \sinh^2\frac{1}{2}\beta t + \frac{\gamma}{\alpha'} \sinh{\alpha' t} + 1)\Big) \Big],\\
    \langle x(t)\rangle &= x_0 e ^{-\frac{1}{2}\beta t} \left( \cosh{\frac{1}{2}\alpha' t} + \frac{\beta}{\alpha'} \sinh{\frac{1}{2}\alpha' t }\right)  + 2\frac{v_0}{\alpha'} e^{-\frac{1}{2}\beta t} \sinh{\frac{1}{2}\alpha't} \\
\end{split}
\end{equation}
where  $\alpha' = \sqrt{\beta^2-4\omega_0^2}$.

The experimental variance displayed in Figure \ref{fig:varianceandMSD}~(a) behaves like an overdamped oscillator, as no oscillations are observed. Note that we have previously removed the linear drift in data by subtracting the average path from each trajectory, so we can use Eq.~\ref{eq:sigmaarmonico} in the overdamped limit to attempt fitting the experimental data. Fitting the results for the experimental variance in $y$ (shown in Figure \ref{fig:varianceandMSD}~(a) in dark gray line), allows obtaining $A = \langle v^2\rangle/\omega_0^2$, $\beta$ and $\alpha$ (see Table \ref{tab:fitparametersConfined}). Notice that, in principle, by using the fact that $\alpha = \sqrt{\beta-8\omega_0^2}$ we could extract the $\omega_0^2$ just from this fitting. However, this procedure would give a huge uncertainty in the resulting parameters. We can however make use of the the values $\langle v^2\rangle$ computed from the sample velocity of each trajectory ($v_i = \Delta x_i/ \Delta t$), in order to determine  $\omega_0^2 = \langle v^2\rangle/A$.

The resulting values are consistent with a strongly overdamped regime. The parameters $\langle v^2\rangle$ and $\beta$ obtained from this procedure can be used to estimate the Reynolds number $R_e\sim 18\langle v\rangle/(a\beta) \sim 2.5$ using Eq.~\ref{eq.reynold}, which means that the inertial term cannot be neglected.\\

\onecolumngrid
\begin{table}[h!]
\begin{minipage}[T]{1\textwidth} 
\centering
\caption{Experimental values obtained from the variance and the sample velocity}
\begin{ruledtabular}
\begin{tabular}{l c c c c c c}
coordinate & 2D [mm$^2$/s] & $A$ [m$^2$] & $\beta$ [1/s]  & $\alpha$ [1/s]  & $\langle v^2\rangle$ [mm$^2$/s$^2$]  & $\omega_0^2$ [1/s$^2$] \\
\hline	  
x &  $16.5\pm 0.1$ &  $162.8\pm 4.8$ & $17.9\pm 1.2$ & $17.7\pm 1.2$ & $181\pm 1$ & $0.91 \pm 0.11$ \\
y &  $17.1\pm 0.1$ &  $173.5\pm 5.0$ & $18.5\pm 1.2$ & $18.3\pm 1,2$ & $183\pm 1$ & $0.95\pm 0.11$ \\
\end{tabular}
\end{ruledtabular}
\label{tab:fitparametersConfined}

\medskip
\end{minipage}
\end{table}
 
The variance was simulated for $\sim$ 500 trajectories using Equation \ref{ec:sim_LMCP} and the parameters obtained from the fitting. It can be seen in Figure \ref{fig:varianceandMSD} (a) that there is a good agreement for variances obtained from the simulated trajectories (red), the experimental trajectories (blue), and the theoretical model for the confined inertial particle (black). 
Figure \ref{fig:varianceandMSD}~(b) shows the MSD for the measured trajectories in grayscale, and the average MSD on the trajectories (black line). As a reference, we have also plotted the theoretical MSD for a Brownian particle ($2Dt$) using the diffusion coefficient from Table \ref{tab:fitparametersConfined} (yellow line). As can be seen in the figure, the measured MSD is smaller than the Brownian theoretical MSD for every time. Consistently with results obtained from the variance, this means that the stochastic motion is subdiffusive. We have also plotted the theoretical MSD for an overdamped particle (green line) computed by using equation \ref{ec:MSDconf} and the coefficients from Table \ref{tab:fitparametersConfined}. Additionally, we simulated 2.5$\times$10$^{6}$ trajectories using the values of Table \ref{tab:fitparametersConfined} and we calculated the MSD for subgroups of 500 trajectories (same number as the experiment). The superposition of all MSD is also plotted in Figure \ref{fig:varianceandMSD}~(b) in the red shadowed area. As can be seen, in agreement with the previous discussion, the measured MSD has a behavior similar to that of the overdamped model. At long times there is a deviation from the model's prediction, but even in this range, the measured MSD has no significant differences from the ones calculated from the simulations. We have observed with the simulations, that taking subgroups of 500 trajectories, the simulated MSD behaves as the measured MSD, meaning that the differences are only a consequence of the statistics sample size (the number of trajectories). In fact, as it can be seen in simulations from supplementary material \cite{supplementary}, as the number of trajectories increases the simulated mean MSD becomes closer to the theoretical MSD.

\subsection{Correlations and Power spectral density (PSD)}
As mentioned before, Brownian motion requires the displacements to be uncorrelated. Then, ideally, the correlation function $\langle \Delta x(\tau)\Delta x(t+\tau)\rangle_{\tau}\propto \delta(t)$. Since both of the displacements are the same coordinate ($x$), the function on the left is known as the \textit{autocorrelation} function. Correlation can also be tested for different coordinates or different trajectories, in which case it is referred to as \textit{cross correlation}, which is expected to be statistically zero in random motion. 

We next compute autocorrelations and cross-correlations for the experimental trajectories and their corresponding averages. Examples of results are shown in the supplementary material \cite{supplementary}. In all the cases, a structured noise that smooths after averaging is observed. Autocorrelations show a high correlation peak at $t = 0$, which rapidly decays to zero in a correlation time around $0.1$ s. In addition, the cross-correlation between displacements from different trajectories and from different coordinates does not show a distinguishable correlation from the background, in agreement with the random hypothesis.

We will now focus on studying the structure of the noise by examining its frequency content. According to the Wiener-Khinchin (WK) theorem,  the power spectrum is the Fourier transform of the correlation function. Therefore, the analysis of the power spectral density (PSD) reveals information about the correlation structure of a signal and how the signal's power or energy is distributed among various frequencies, as discussed by Krapf\cite{Krapf}. By examining the Fourier transform of correlations, we can gather information about the fluctuation spectra and, consequently, the type of noise associated with the studied process.

The power spectrum is defined as
\begin{equation}
\begin{split}
    S_{\Delta x }(\omega)&=\mathfrak{F}\{\langle \Delta x(\tau)\Delta x(t+\tau)\rangle_{\tau}\}(\omega)\\ 
   &= \int_{-\infty}^{\infty} dt \, e^{-i\omega t} \langle \Delta x(\tau)\Delta x(t+\tau)\rangle_{\tau}.
    \label{ec:PS}
\end{split} 
\end{equation}

Therefore, if the random condition is fulfilled, a constant PSD (white noise) is expected.

Since the displacements are related to the velocity by a constant increment of time ($\Delta x_i = v_i \Delta t$), the statistics for the displacements and velocities are similar. However, for an inertial particle subjected to a trapping potential in a thermal media, it can be shown that the PSD of velocities ($S_{v}$) can be expressed as follows\cite{PhysRevE.83.041103} 

\begin{equation}
\begin{split}
    S_{v}(\omega) &=\mathfrak{F}\{\langle v(t) v(t+\tau)\rangle_{\tau}\}(\omega)\\ 
        &=\frac{A_{v}}{2\pi}  \cdot \frac{\omega^2}{(\omega_0^2-\omega^2)^2 + \beta^2 \omega^2},
  \label{ec:SvConfined}
\end{split}
\end{equation}
where $A_{v}$ is a normalization constant. 
  
\begin{figure}[htp]
    \centering
    \includegraphics[scale=0.4]{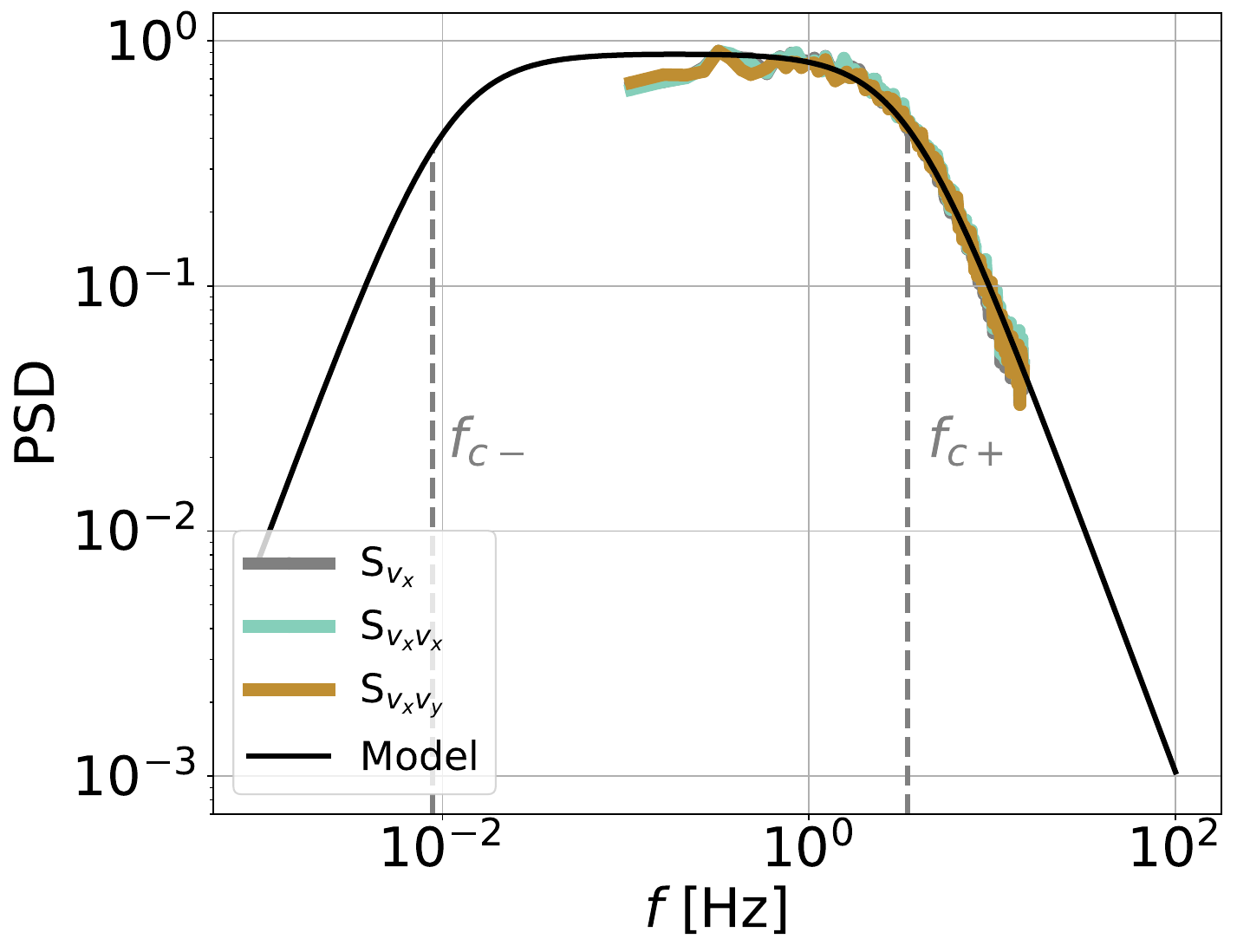} 
    \caption{Normalized power spectral density of the velocities calculated from the autocorrelation of displacements in x ($S_{v_x}$, gray line), from the cross-correlation of the displacement in x in different trajectories ($S_{v_x v_x}$, cyan line) and from the displacements in x and y  ($S_{v_x v_y}$, brown line). The black line is the spectral density expected in an overdamped regime, following the expression in Equation \ref{ec:SvConfined}. The critical frequencies separate the confined, Brownian, and inertial regimes}
    \label{fig:PSD}
\end{figure}

Normalized experimental PSDs, associated with correlations of velocities were obtained by estimating the average PSD from the correlation of displacements normalized to their maximum. Due to limitations imposed by temporal and spatial sampling, experimental results cannot be fitted directly by the expression in Equation \ref{ec:SvConfined} (see supplementary material \cite{supplementary})\cite{PhysRevE.83.041103, oe-18-8-7670, 10.1063@1.1645654}. In terms of data analysis, it is very interesting to discuss with students the consequences arising from limitations of detection and numerical computation. However, due to the complexity of this type of analysis, we can only recommend that the most dedicated students pursue a genuine path in this direction, if they have enough time.  Among others, the errors can be related to the numerical approximation used in the velocity, the integration time of the camera, the choice of the sampling frequency, and the error in the measured position (see supplementary material \cite{supplementary}). To be able to fit the normalized PSD for the displacements $S_{\Delta x}$ with the expression in Equation \ref{ec:SvConfined}, we need to add a constant parameter that, in this particular case, resulted in $C \sim -7\times 10^{-2}$. This constant was then added to the data, in order to obtain $S_v = S_{\Delta x}+ C$. Figure \ref{fig:PSD} shows the average PSDs for velocities (color curves), obtained after adding the $C$ constant, to either correlations or cross-correlations of displacements. As can be seen, all the PSDs look similar. 
 
The theoretical curve predicted by Equation \ref{ec:SvConfined}, with the parameters shown in Table \ref{tab:fitparametersPSD} is also shown with a black line. A very good agreement between the experimental data and the model for an overdamped regime is obtained. Moreover, this data could be well represented with parameters very close (in some cases coincident) to those reported in  Table \ref{tab:fitparametersConfined}.  

\begin{table}[h!]
\centering
\caption{Parameters obtained by fitting the PSD}
\begin{ruledtabular}
\begin{tabular}{l c c}
coordinate &  $\beta$ [1/s]  &  $\omega_0^2$ [1/s$^2$] \\
\hline	 
x &  $17.9\pm 1.2$ &   $0.91 \pm 0.11$ \\
y &  $21.6\pm 0.4$ &   $1.1 \pm 1.5$ \\
\end{tabular}
\end{ruledtabular}
\label{tab:fitparametersPSD}
\end{table}

From Equation \ref{ec:SvConfined} we can define two corner angular frequencies:~$\omega_{c\pm} = \pm \beta/2 + \sqrt{(\beta/2)^2 + \omega_0^2}$, and the corresponding corner frequencies $f_{c\pm}=\omega_{c\pm}/2\pi$. The lowest corner frequency separates the confined and unconfined Brownian regimes in a potential represented by a simple harmonic oscillator and the higher one determines where the inertial term cannot be neglected. The corresponding corner frequencies $f_{c+}$ and $f_{c-}$ are indicated by dashed lines in figure \ref{fig:PSD}. For frequencies in between  $S_{v}$ tend to be constant, so the trajectories would be similar to a Brownian motion. For the highest (lowest) frequencies the power spectrum decreases (increases) linearly in the logarithmic scale with a slope of $-2$ ($+2$). It can be seen that our data are far away from the lowest corner frequency $f_{c-}$, indicating that confinement should not be too relevant and could be neglected under this model. The subdiffusive observed behavior should therefore imply that the actual confinement potential may not be modeled for a harmonic oscillator at long times. 

On the other hand, the highest corner frequency $f_{c+}$ ($ \sim 2.8-3.5 \, Hz $) is included in the experimental frequency range. This means that for sample times less than $\sim  0.3 \, s$, correlations due to inertial effects can be noticed and displacements are still not totally diffused by viscosity and noise. This effect is also observed in the direct computations of correlations, which deviate from zero within a time span of 0.2 s (see supplementary material\cite{supplementary}). 
If inertia is not negligible (meaning the particle continues to accelerate during a measurable time while undergoing impacts), a bias will be introduced into the trajectory, resulting in correlations that are distinct from zero. Therefore, these results are in agreement with the previous conclusions from the estimated Reynolds number, indicating that the inertial term cannot be neglected.

\section{Closing Remarks}\label{sec:conclusions}

The analysis of stochastic processes focused on Langevin dynamics generally requires several tools and physical concepts, usually not taught in undergraduate courses. The present article proposes a simple experiment, aiming for undergraduate students to become familiar with concepts and statistical approaches that allow them to embrace this kind of system.

The experiment was proposed by professors in the advanced undergraduate Laboratory course (Laboratorio 5) taught in the Department of Physics, at the Faculty of Exacts and Natural Sciences, University of Buenos Aires.  The general setup has been developed and built in collaboration with technical staff from the teaching laboratories. The experiment generally takes three classes of 6 hours each separated a week apart, so students have time in between to write the software for the analysis. We don't encourage university professors to provide ``recipes" to follow when conducting an experiment. Instead, we prefer to guide the students through the discussions that emerge in the laboratory as they perform the experiments.  After the three classes they write a report, so they can also have time to complete the analysis and discuss the results with the teachers.  Some necessary general statistical concepts were already developed in the first weeks of the same course. Before the first experimental class, students are asked to read the material and the associated topics, both regarding basic physics concepts (in this case, an introduction to stochastic trajectories, Langevin dynamics, and Brownian motion) and about the experimental setup they will be using. This basic information together with complementary bibliography is provided online (see ref. \citenum{pagina}). Results presented in this paper were obtained by a group of three students, which performed all the data acquisition and analysis, under the supervision and assistance of the teacher team.

The main conclusions come from the computation of the variance and the MSD of the analyzed trajectories, which allows detecting, that, in this particular experiment, the particles move in a sub-diffusive regime.  In order to model a more comprehensive system that includes the inertial term and a potential confinement condition, a restorative force was incorporated into the Langevin equation. This adjustment allowed for the fitting of the experimental variance using the model proposed by Lemons, resulting in the acquisition of relevant parameters. These parameters were subsequently used to conduct numerical simulations, whose outcomes demonstrate strong agreement with both model predictions and experimental results.

The obtained parameters allowed us to estimate a Reynolds number of around 2.5, indicating that, in this particular experiment, the inertial term cannot be neglected.

The random hypothesis was also tested using correlations, which led to correlation times of around 0.1~seconds. Finally, the power spectrum density (PSD) of displacements (or velocities) was analyzed and compared with the PSD expected from the full model, resulting in a good agreement throughout the entire frequency range covered by the experiment. Although the PSD calculated from autocorrelation and cross-correlation may not be expected to be necessarily similar, we observe that in this case, they behave almost identically. We believe this is because the correlation length is small, and the nature of the noise is the same in all the involved trajectories.

The experimental frequency range includes the higher corner frequency, confirming that the inertial term cannot be neglected. Moreover, the lower experimental frequencies are far away from the lowest corner frequency, indicating that confinement should not be relevant in the proposed model. The subdiffusive behavior, observed at times longer than 2 seconds, therefore suggests that the harmonic potential is not a very good representation of the real confinement potential in the large time scale, and higher terms in the potential approximation should be included to reproduce these results. 

We remark on additional important concepts to discuss with students, related to the limitations arising from sampling frequency and limited range of time and space, and finite statistical sample size (number of trajectories). While a quantitative study of these limitations surely exceeds the possibilities of a regular undergraduate course, we consider that it is important to mention some of them. For example, in order to obtain a good representation of the expected MSD, the statistical sample size needs to be of the order of 5000, 10 $\times$ larger than the actual number of measured trajectories (see supplementary material \cite{supplementary}). Other quantities such as variance, PSD, or correlations do not require such a large number of measurements. 

Additionally, we propose a complementary analysis, based on the comparison of the experimental probability distribution with the Gaussian distribution, in order to contrast with that expected from the hypothesis of a simple Brownian motion. In doing so, we used chi-square tests and (Q-Q) plots.

We would also like to emphasize that the results obtained in this work correspond only to one of the possible configurations for the experimental setup. However, there is a great playground for exploration in terms of plate oscillations, by testing different modes combinations. Some questions that can guide the students to perform other experiments are: How can you make the Brownian trajectories longer? What happens to trajectories if the plate oscillates in a mode that has a node in its center? Would it be beneficial to change the plate for another with a larger diameter? Can a square or triangular wave be used to produce the plate excitation? How many modes are important in that case? Do variations in the weights and sizes of the beads lead to changes in the experimental results?

In supplementary materials online, we included additional information that complements the main content of this document. We included resources for those interested in replicating or extending this study, such as phyton codes (Tracking algorithms, MSD simulation, software to control the speaker), several proposals to further analyze the data (study of the deviation from a Gaussian distribution, influence of sampling of time and space and number of trajectories in the PSD and MSD), descriptions and tutorial for the tracking software.

In summary, we present a simple experiment that has been tested to be implemented in undergraduate courses. This experiment provides a practical way to introduce several complex tools and concepts. Moreover, we demonstrate that a rich statistical analysis of stochastic trajectories can be performed, allowing the students to obtain information about the underlying microscopic physics.

\begin{acknowledgments}
The authors acknowledge useful discussions and insights during the experiment development from H. Grecco, G. Patterson, and D. Shalom, as well as further discussions regarding models and numerical procedures with G. Lozano, L. Cugliandolo, Julian Fernandez Bonder, and Pablo Fierens. This work was supported by Universidad de Buenos Aires, Facultad de Ciencias Exactas y Naturales, and Departamento de F\'isica.
\end{acknowledgments}
\vspace{1em}
The authors have no conflicts to disclose




\end{document}


\title{Supplementary material: Tracking algorithm}
\maketitle

\section{The algorithm}

\begin{figure}[htp]
    \centering
    \includegraphics[scale=0.15]{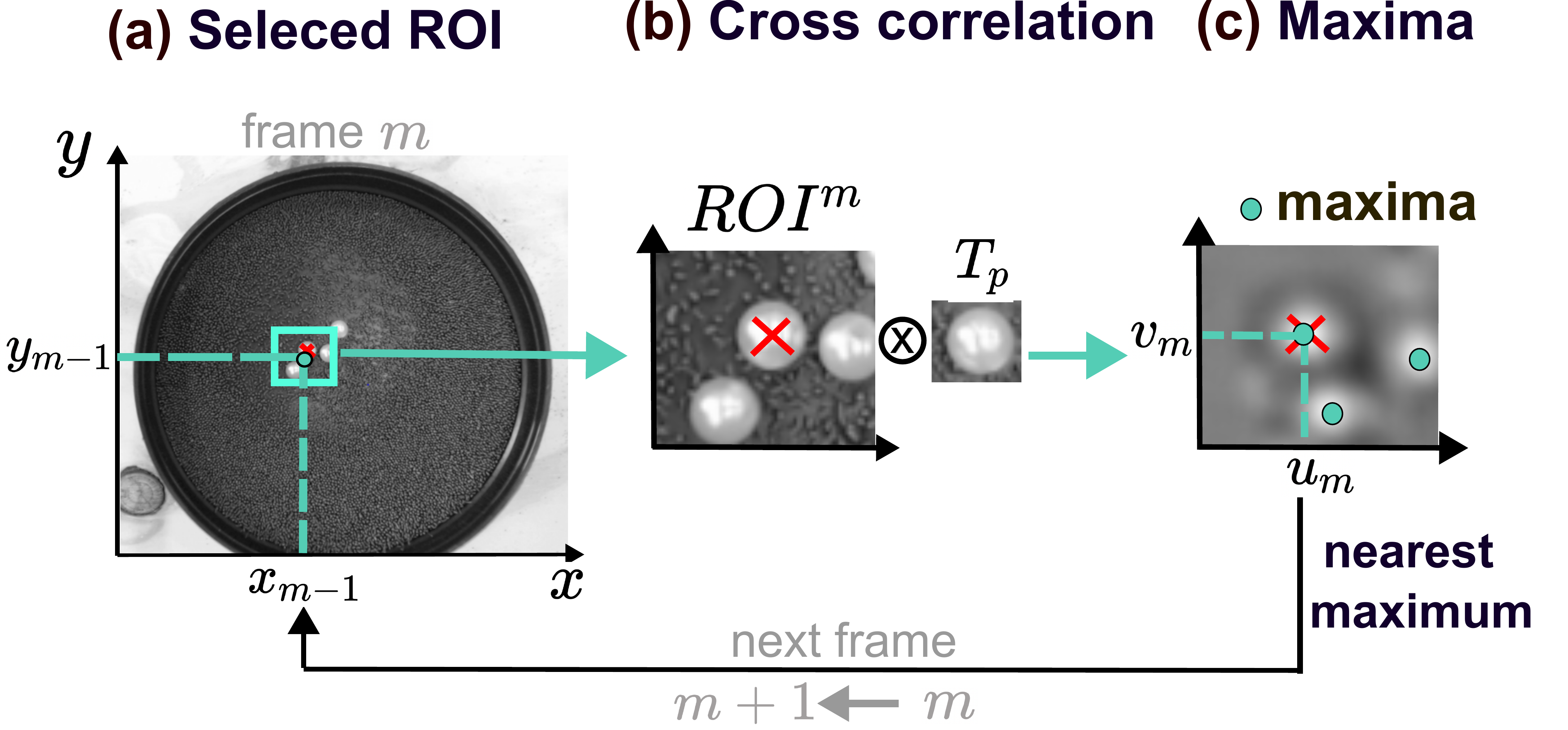} 
    \caption{Algorithm's flow diagram: (a) Selects $ROI^m$ in the frame $m$ centered at the pixel $(x_{m-1}, y_{m-1})$. (b) Computes the cross-correlation between $T_P$ and $ROI^m$. (c) Determines the maxima and chose the nearest position to $(x_{m-1}, y_{m-1})$, that is $(x_{m}, y_{m})$ and shift ROI to that position. (d) advance one step to the next frame  $(m+1)$}
    \label{fig:tracker}
\end{figure}

As a means of capitalizing on the experiment for teaching various tools in image processing, students developed a program based on correlations to track each particle. A flow diagram of the tracking code is shown in Figure \ref{fig:tracker}. The general idea is to find, in each frame, a region similar to a template $T_P$, which is an image of one plastic bead. This template is selected from the first frame and is kept during the whole evolution of the trajectory. We then search for the region in each frame that maximizes the cross-correlation and record its position. The initial position $(x_o, y_o)$ of the bead center serves as the starting point for the trajectory. In practice,  to shorten the computations times (which are typically large for cross-correlations calculations), we then select a region of interest (ROI) centered in $(x_o, y_o)$, which corresponds to the image portion where the white beads are located. When choosing the size of the ROI, it is important to consider an estimation of the average size of the displacements. This is because the displacement from one frame to the next needs to be smaller than the ROI in order to keep tracking. In other words, a ``single jump" between frames should be smaller than the ROI. 

The algorithm works as follows: let'
s assume it is in the iteration step $m$ and the particle has been located in the previous step at the position $(x_{m-1}, y_{m-1})$. Then, for this step, the $ROI^m=ROI(i,j,m)$ is initially centered at that position. The best match is found by maximizing the modulus of the cross-correlation between the ROI and the template, which is essentially a matrix given by the inner product between the template and the 2D shifted version of the ROI, with $(u,v)$ the displacement vector. The definition for the cross-correlation we used is

\onecolumngrid
\begin{equation}
    C(u,v,m)=\sum_{i=-N_T/2}^{N_T/2}
    \sum_{j=-N_T/2}^{N_T/2} [(T_P (i+N_T/2,j+N_T/2)-\overline{T}_P)
    \cdot(ROI^m(i+u,j+v,m)-\overline{ROI}^m)],\\
\end{equation}

where $\overline{T}_P$ and $\overline{ROI}^m$ are the average values over the pixels of $T_P$ and $ROI$ respectively, and $N_T$ is the size of the template in both dimensions.  The displacements $(u,v)$ are relative to the ${ROI}^m$ central position, and take values in $[-(N_R+N_T-1)/2, (N_R+N_T-1)/2]$, where $N_R$ is the size of the ROI.

If the maxima of $C(u,v, m)$ is located at  $(u_m, v_m)$, the positions of the bead in the step $m$ is $(x_m, y_m)=(x_{m-1}+u_m, y_{m-1}+v_m)$. If many beads are in the ROI, to define the trajectory of one bead in particular, the algorithm chooses the position $(x_m, y_m)$ that is closer to the position of this bead in the previous step (that is $(x_{m-1}, y_{m-1})$). Then a new ROI for the frame $(m+1)$ is selected ($ROI^{m+1}$), with the same size as the previous but centered it $(x_m, y_m)$. Then the algorithm advances to the step $(m+1)$. In this way, we can find where the object has moved between frames. The algorithm continues updating the area of observation so it follows the particle and no further input is needed.

\section{Step-by-Step: Guide to Using the Tracker Program}
The Python script consists of two files: `trackerclass.py' and `take\_data.py'. The first one is a module that contains functions for selecting the template and creating the tracker, while the second file uses these functions to specify the parameters and execute the tracker. The script is available for reference in gitHub repository \cite{github} and can be implemented by users with basic Python knowledge. As visual help, we have also uploaded a video demonstrating how to use it \cite{youtube3}.

\begin{figure}[h]
    \centering
    \includegraphics[scale=0.18]{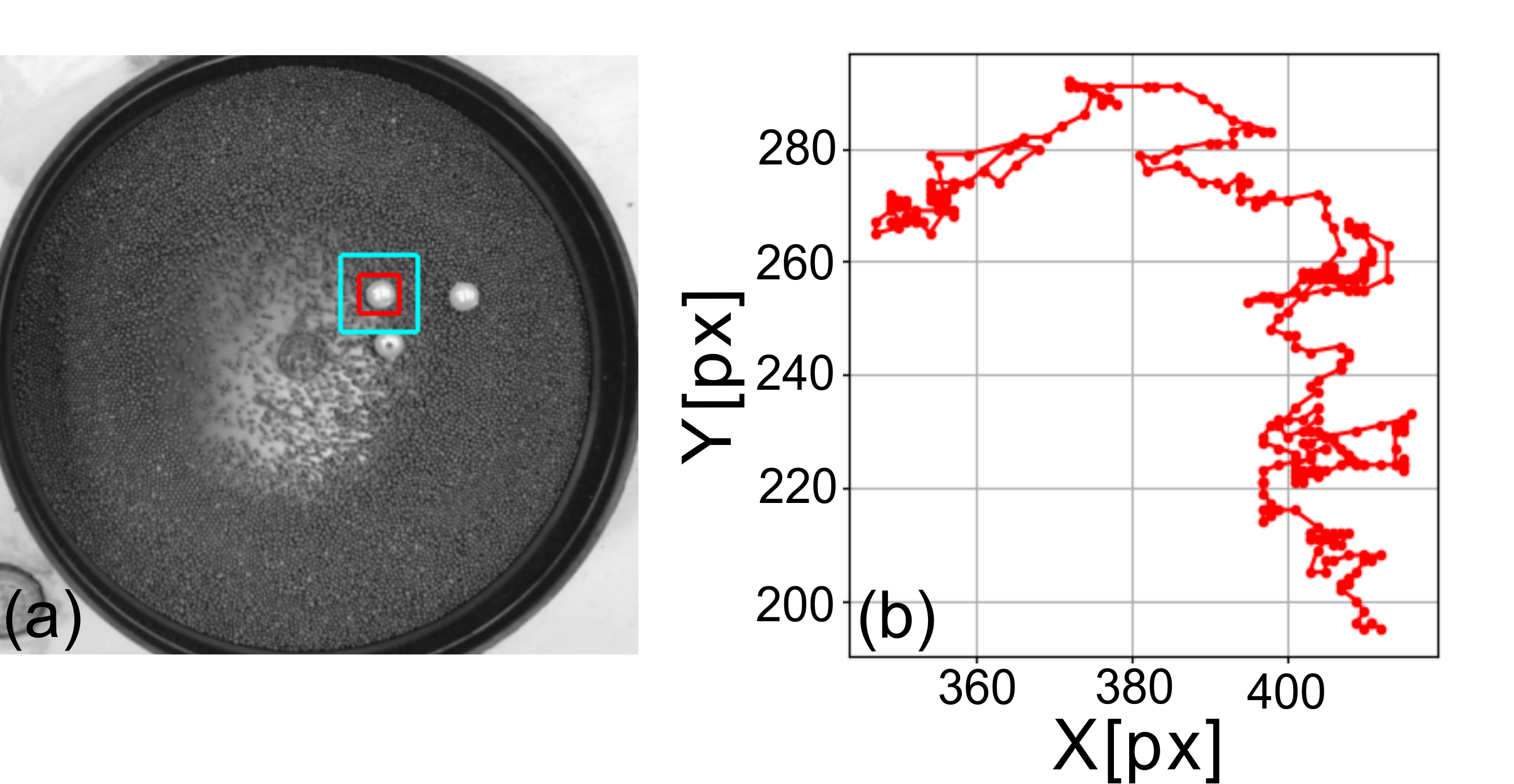} 
    \caption{Tracker program instances. (a) Step 4. Particle tracking. (b) Step 5. Review of tracking results.}
    \label{fig:app_tracker_program}
\end{figure}

Before starting, ensure that the following libraries are installed on your operating system: `numpy', `matplotlib', `openCv' and `scikit-image'. These libraries are necessary for the program to run properly. Once this is well done, is possible to execute the tracking with the following steps:
\begin{itemize}
    \item[\textbf{1}]\textbf{Specify the video file and initial parameters:} Open the `take\_data.py' file and assign the value 
     of the video file name to the variable `videoName'. Ensure that the video file is in the same folder as the script. Choose the initial time and duration  of the tracker and write them into the variables with the same name. The height and width of the Region of Interest (ROI) can be settled into the attribute `observationWidth'. Could be convenient to make this region smaller when the particle is too close to another one so the tracker doesn't  mix the target.
     
    \item[\textbf{2}]\textbf{Execute the module:} Run the `take data' module on Python by typing `python take\_data.py' in the command prompt or terminal. Make sure that the `trackerclass.py' file is also present in the same folder as the `take data' file.
    
    \item[\textbf{3}]\textbf{Select Template:}  After executing the module, an emergent window will appear. Use the rectangle tool to select the particle you want to track. Press and hold the left mouse button to draw a rectangle around the particle. Once you have selected the particle, press the `ENTER' key to define the template. If you want to cancel the selection process, press the `c' key.
    \item[\textbf{4}]\textbf{Particle tracking:} After defining the template, the script will start tracking the particle automatically in the window. The particle's position will be marked with a red rectangle, while the ROI will be marked with a cyan rectangle (Figure \ref{fig:app_tracker_program}~(a)). If the particle is too close to another one, you can adjust the size of the ROI to make it easier for the tracker to follow the objective.
    \item[\textbf{5}]\textbf{Review of the tracking results:} Once the tracking is complete, a graph of the trajectory will be plotted automatically, as is shown in the Figure \ref{fig:app_tracker_program}~(b). The tracking results will be saved in the `xcorr' and `ycorr' variables, which contain the x and y coordinates of the particle at each time step and in pixel units. 
\end{itemize}

\vspace{3em}




\title{Supplementary material: Additional data analysis}
\maketitle

\section{Correlations} \label{ap:Correlations}

\begin{figure}[h]
    \centering
    \includegraphics[scale=0.32]{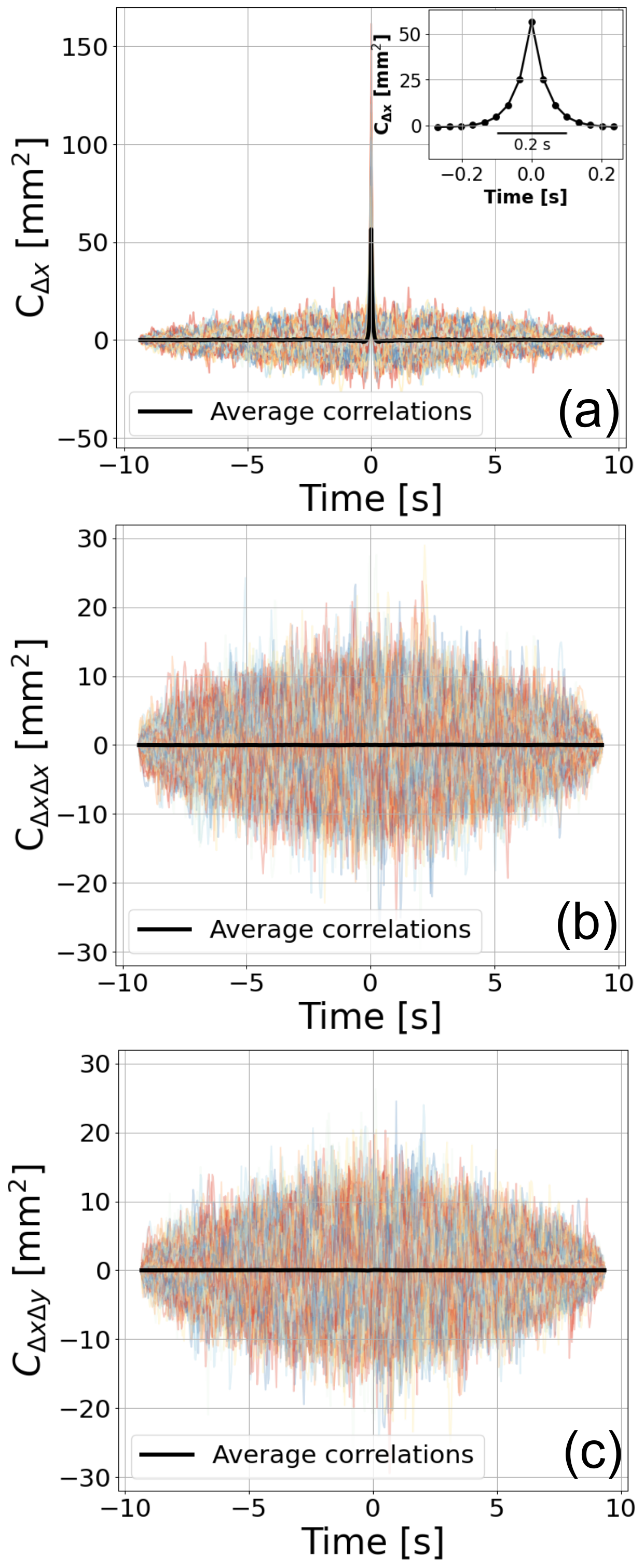} 
    \caption{(a) Autocorrelation and (b) Cross correlation for  x displacements. (c) Cross-correlation between x and y displacements. Average correlations are plotted in black. }
    \label{fig:Correlations}
\end{figure}

To corroborate that collisions produce uncorrelated random displacements, autocorrelations and cross-correlations for the trajectories together with their corresponding averages were computed. Examples of results are shown in Figure \ref{fig:Correlations}. Autocorrelations (Figure \ref{fig:Correlations}~(a)) show a high correlation peak at $t = 0$, which rapidly decays to zero in 3$\times \Delta t \sim 0.1$ s, meaning that there is a characteristic time (correlation time) where the displacements are correlated (see inset in Figure \ref{fig:Correlations}~(a)).  The cross correlation between displacements from different trajectories and from different coordinates are shown in Figure \ref{fig:Correlations}~(b) and (c) respectively. None of them show a distinguishable correlation from the background. It's noticeable as well that the mean value moves around zero.

\section{Beyond the Gaussian distribution}\label{ap:qqplot}

As was discussed in Section IV-A (main text), the moments of the distribution give estimators for the mean trajectory $\bar{x}= E[x]$ and the variance $E[(x-\bar{x})^2]=\langle (x-\bar{x})^2 \rangle = \sigma^2(t)$. The third and fourth moments are related to skewness and kurtosis respectively.  The skewness ($\mu_3/\sigma^3$) characterizes the symmetry of a distribution and the kurtosis ($\mu_4/\sigma^4$) characterizes how heavy tail the distribution is. Both moments are zero for a Gaussian distribution. 

As discussed in the main text, Equation 9 (main text) predicts a Gaussian distribution for the positions of a Brownian particle. Therefore, an alternative or complementary way to test for a deviation from Brownian dynamics consists of exploring the deviation of this prediction. In our case, to study if this prediction is accomplished, as a first approach, we plot together with each histogram a Gaussian function using the mean value and standard deviation calculated from the data (blue continuous lines in Figure 4 (main text). At first sight, there is an apparent accordance between the histograms and the Gaussian functions. However, after estimating the Chi-squared and the corresponding p-values (between 0.02 and 0.05), we conclude that the certainties of the Gaussian overlap don't reject the null hypothesis.

A simple but powerful tool to analyze if a Gaussian model matches with a given distribution is the  Quantile-Quantile plot or Q-Q plot. Briefly, this method compares the quantiles for the data with the quantiles for the expected Gaussian distribution: if a Gaussian model holds, a linear behavior is expected in the Q-Q plot.

Quantiles are points that divide a data set into equally sized groups. For example, the median is a quantile because it splits the data set into two groups containing the same number of data points. This quantile is labeled as 0.5 or 50$\%$. If we now divide the data set into four equally sized groups, we are adding the 0.25 and 0.75 quantiles. The total number of quantiles is one less than the number of groups into which the data is partitioned. In this framework, Q-Q plot is a simple method that compares the quantiles from the data (generally plotted on the y-axis) and the quantiles for the expected Gaussian distribution (in the x-axis). The first step to creating the Q-Q plot is to assign each point in the data its corresponding quantile, known as the sample quantile (SQ). Next, we generate the theoretical quantiles (TQ) by calculating an equal number of quantiles for the expected normal distribution. Finally, we plot the sample quantiles (SQ) against the theoretical quantiles.

\begin{figure}[htp]
    \centering
    \includegraphics[scale=0.3]{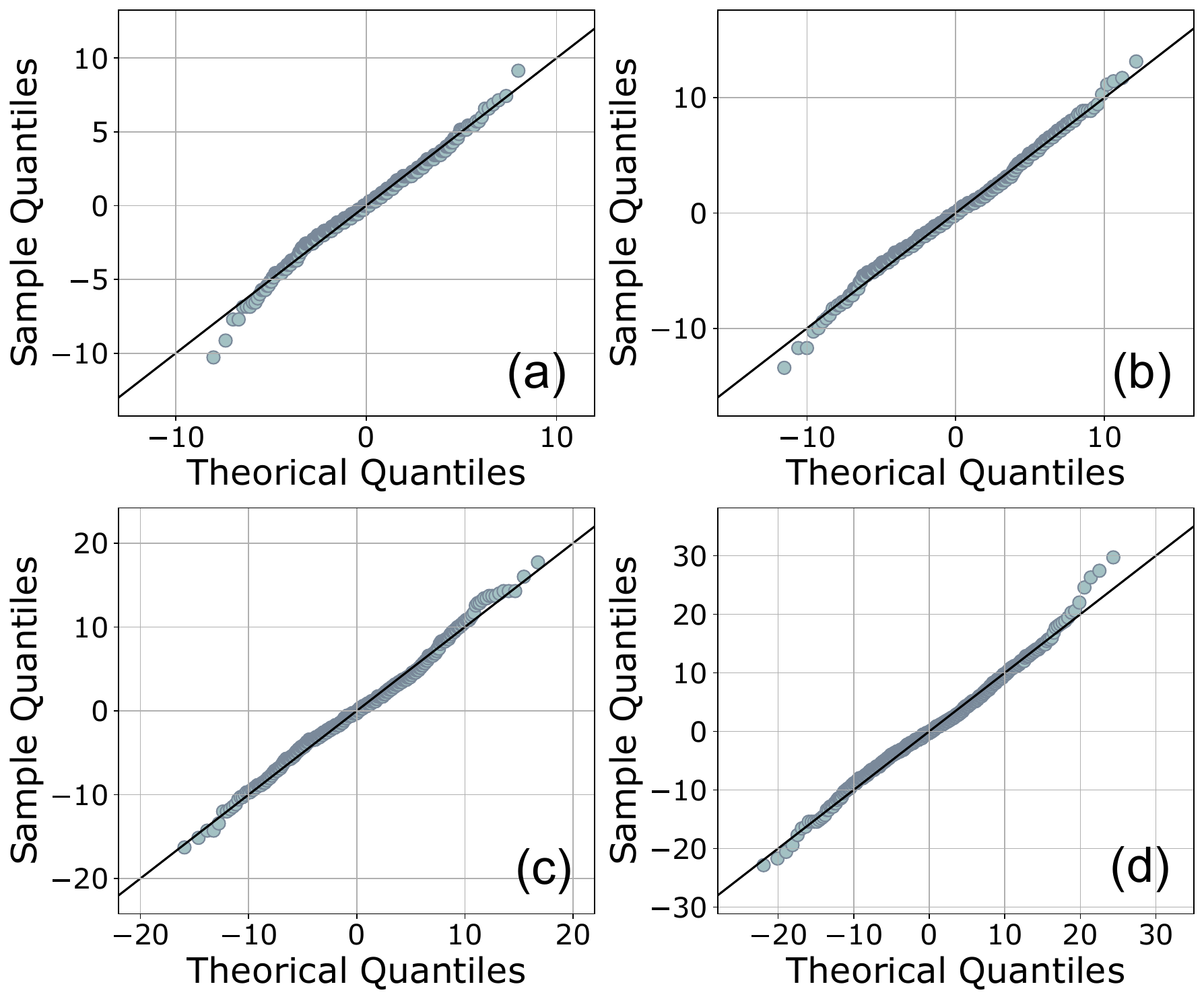} 
    \caption{(Q-Q) plots for Gaussian distributions corresponding to histograms shown in Figure 4 (main text), where TQ is for theoretical quantiles and SQ is for sample quantiles.}
    \label{fig:qqplots}
\end{figure}

The Q-Q plots comparing each histogram with its Gaussian distribution are shown in Figure \ref{fig:qqplots}. At the beginning of the movement (short times, panel (a)), the Q-Q plots have a slight deviation from the straight line in the bottom end; therefore the experimental distribution has a longer tail to its left than the normal distribution (i.e. it is left-skewed). On the other hand, at intermediate times (panels  (b) and (c)) the quantiles spread on the line, so the Gaussian assumption matches with the data. For larger times (panel  (d)), the experimental distribution shows a deviation in the upper end (right-skewed). The skewness and kurtosis estimated from the distribution moments resulted in the order of  $10^{-5}$. Even though these values are very small, the Q-Q plot makes the asymmetries evident. 

\begin{figure}[htp]
    \centering
    \includegraphics[scale=0.3]{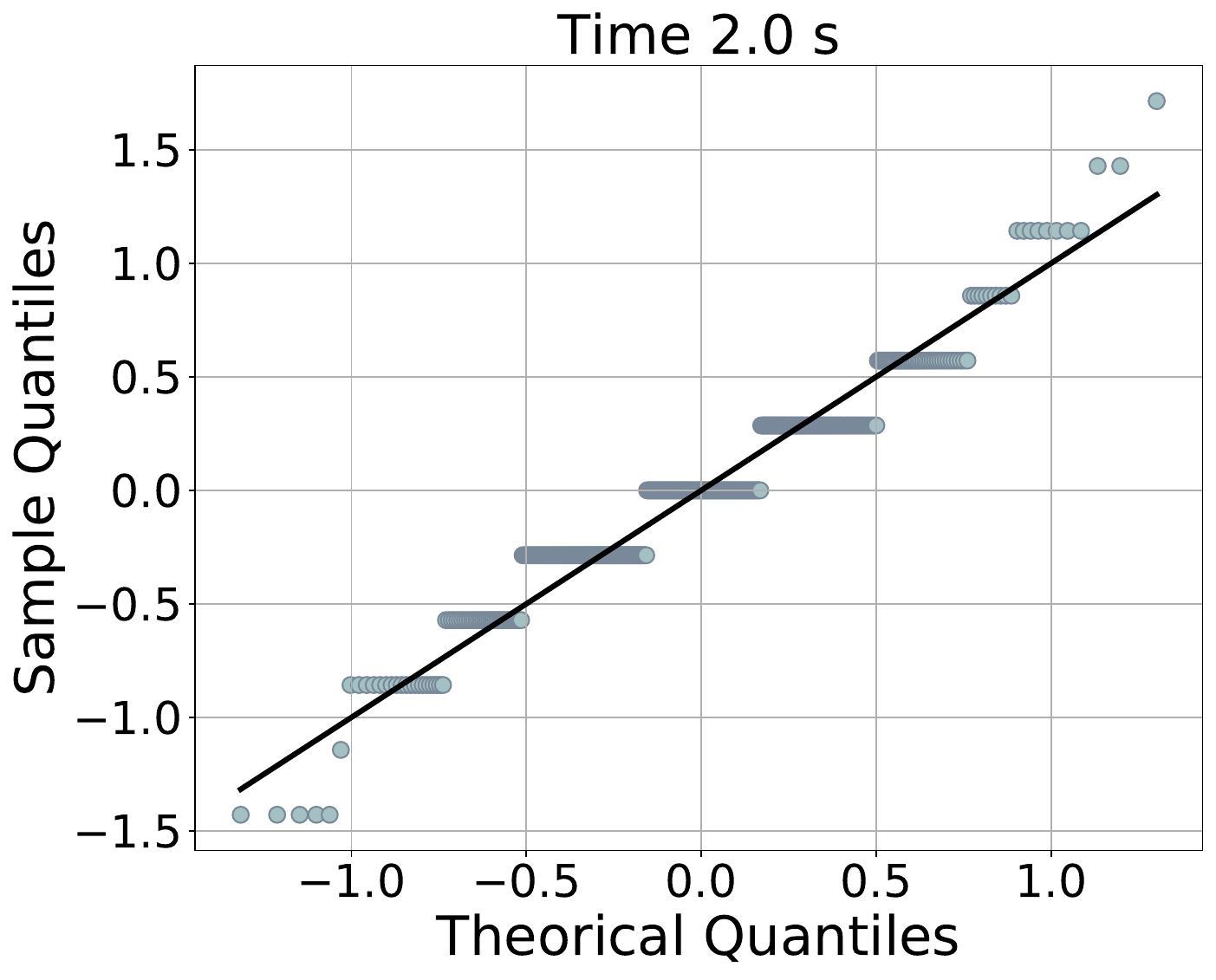} 
    \caption{Q - Q plot for displacements in coordinate x at 2 s.}
    \label{fig:displacementQQplot}
\end{figure}

A similar analysis can be conducted with the displacements, which are calculated as the differences between consecutive positions: $\Delta x_{n} = x_{n+1} - x_{n}$. The construction and predictions of Brownian trajectories can be developed for the displacement $\Delta x$ using Equation 5 (main text), leading to a Gaussian probability density function centered at a deterministic step. Figure \ref{fig:displacementQQplot} shows the Q-Q plot for the displacements at 2 seconds (Q-Q plots for different times are similar). The observed stepped dependency is caused by the discretization of the trajectories that are acquired by a multiple-pixel detector with an inter-pixel distance of ($ \sim 0.28$)~mm, which determines the minimum detectable displacement. The quantiles of the displacement distribution spread in a line but with a slight deviation at both ends, relating the data with a normal distribution with fat tails (fat-tailed distribution). 

\section{Additional tools to explore} \label{ap:MSD}
In this appendix, we include some additional tools related to limitations in numerical simulations and/or discrete detection that can be explored with students in case they have the necessary background and there is enough time. Otherwise, we recommend mentioning the existence of these factors without delving into them in detail.

\subsection{Sample size in MSD simulations}
\begin{figure}[htp]
    \centering
    \includegraphics[scale=0.3]{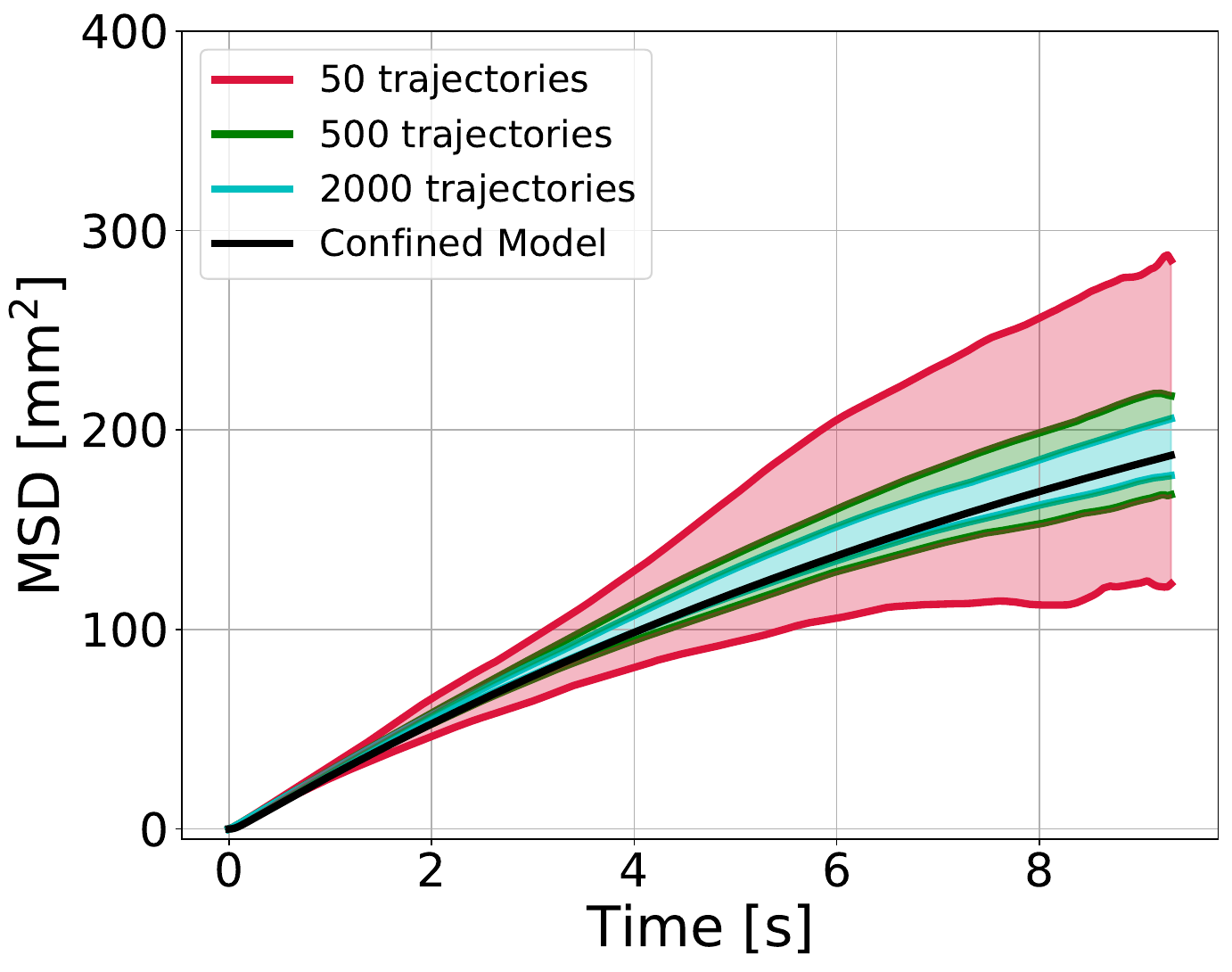} 
    \caption{The colored
area in the plot is the superposition of all the mean MSD calculated from 500 simulations for different sets of 50, 500, and 2000 trajectories.}
    \label{fig:MSD_simulation}
\end{figure}

Figure \ref{fig:MSD_simulation} shows the influence of the statistical sample size on the probability of obtaining an MSD close to the theoretical one. Simulations were made by calculating 500 sets of 50, 500, and 2000 trajectories. The script is available for reference in GitHub repository and can be implemented by users with basic Python knowledge \cite{github2}. The colored area in the plot is the superposition of all the MSD for each set size. When the size of the set increases, it's much more probable that the MSD gets closer to that predicted by the model. 

\subsection{Limitations and errors in PSD computation}

Here, we provide a brief summary of some of the causes of errors in PSD. As shown in refs [\citenum{PhysRevE.83.041103, oe-18-8-7670, 10.1063@1.1645654}], several factors affect the proper measurement of the PSD: first, we are making an approximation for the velocity, assuming that $v_i$ is a secant velocity. To correct this error, the expression for $S_v$  should be modified with a multiplicative factor that depends on the $\mathrm{sin}^2(\pi \omega/2 \omega_s)$ \cite{PhysRevE.83.041103}. Second, the integration time of the camera $\tau_{int}$ might blur higher frequencies in the experiment. It has been shown that this will produce a modulation $\mathrm{sinc}^2(\tau_{int}\omega /2)$ of the spectrum. Thirst, due to the Nyquist theorem, the sampling frequency $\omega_s$ should be larger than $2 \omega_{max}$, the maximum frequency contained in the spectrum, otherwise introducing alias frequencies ($\omega+n \omega_s$) in the spectrum. Last and most important, the position error of the particle, independent of the frequency, will introduce a constant noise level in the detection, causing a possible offset between the measured $S_{meas}$ and the underlying $S_v$. Therefore, to be able to fit the obtained PSD with the theoretical model, the simplest correction consists of including a constant in the fitting procedure.

